%% file: main.tex
\documentclass[acmsmall,screen]{acmart}
\usepackage{xspace}
\usepackage{subcaption}
\usepackage{xcolor}
\usepackage{tikz}
\usepackage{enumitem}
\setlist[itemize]{noitemsep, topsep=0pt}
\setlist[enumerate]{noitemsep, topsep=0pt}
\usepackage{circledsteps}
\usepackage{tabularx}
\usepackage{multirow}

\AtBeginDocument{%
  }

\acmDOI{XXXXXXX.XXXXXXX}

\titlenote{This paper is a substantial extension of our previous work accepted at SOSP 2024 under the same title. In this journal version, we focus on LLM agents as an emerging serverless workload and provide a comprehensive workload characterization and cost analysis. We introduce new techniques such as browser sharing, extend \systemname to support VM-based deployments, and redesign memory sharing mechanisms using virtio-pmem. We also modify the Cloud Hypervisor restore path to support our design. Finally, we significantly expand the evaluation, demonstrating \systemname's advantages over state-of-the-art systems like E2B.}

\begin{document}

\title{\update{\systemname: Transparently Share Serverless Execution Environments Across Different Functions and Nodes}}

\author{Jialiang Huang}
\email{huangjl22@mails.tsinghua.edu.cn}
\orcid{0009-0001-7413-3362}
\affiliation{%
  \institution{Tsinghua University and Alibaba Group}
  \city{Beijing}
  \country{China}
}

\author{Teng Ma}
\orcid{0000-0002-7104-1526}
\affiliation{%
  \institution{Alibaba Group}
  \city{Beijing}
  \country{China}}
\email{sima.mt@alibaba-inc.com}

\author{Zheng Liu}
\orcid{0000-0002-0911-267X}
\affiliation{%
  \institution{Alibaba Group and Zhejiang University}
  \city{Beijing}
  \country{China}}

\author{Sixing Lin}
\orcid{0009-0000-8796-5136}

\author{Kang Chen}
\orcid{0000-0002-8368-1109}
\author{Jinlei Jiang}
\orcid{0000-0003-4034-7490}
\author{Xia Liao}
\orcid{0000-0001-6823-758X}
\author{Yingdi Shan}
\orcid{0009-0001-5019-8305}
\author{Yongwei Wu}
\orcid{0000-0002-6651-7032}
\affiliation{%
  \institution{Tsinghua University}
  \city{Beijing}
  \country{China}}

\author{Ning Zhang}
\orcid{0009-0000-5102-5601}
\author{Mengting Lu}
\orcid{0000-0003-2333-2123}
\author{Tao Ma}
\orcid{0009-0008-8629-6274}
\affiliation{%
  \institution{Alibaba Group}
  \city{Beijing}
  \country{China}}

\author{Haifeng Gong}
\orcid{0000-0002-0359-2289}
\affiliation{%
  \institution{Intel}
  \city{Beijing}
  \country{China}}

\author{Mingxing Zhang}
\orcid{0000-0001-7518-0753}
\affiliation{%
  \institution{Tsinghua University}
  \city{Beijing}
  \country{China}}
\email{zhang\_mingxing@mail.tsinghua.edu.cn}







\renewcommand{\shortauthors}{Jialiang et al.}

\newcommand{\systemname}{\textsc{TrEnv-X}\xspace}

\newcommand{\dummyfig}[3]{
  \centering
  \fbox{
    \begin{minipage}[c][#2\textheight][c]{#3\textwidth}
      \centering{#1}
    \end{minipage}
  }
}

\newcommand{\dummytab}[1]{
\begin{minipage}{#1\textwidth}
    \centering
    \begin{tabular}{rcl}
    \toprule
        right & center & left \\
        right & center & left \\
    \bottomrule
    \end{tabular}
\end{minipage}
}

\newcommand{\newc}[1]{#1}
\newcommand{\update}[1]{#1}

\begin{abstract}
Serverless computing is renowned for its computation elasticity, yet its full potential is often constrained by the requirement for functions to operate within local and dedicated background environments, resulting in limited memory elasticity. 
To address this limitation, this paper introduces \systemname, a co-designed integration of the serverless platform with the operating system and CXL/RDMA-based remote memory pools. \systemname's core innovations are repurposable sandboxes, which can be shared across different functions to decrease the associated creation overhead, and OS-level memory templates, which enable rapid state restoration from CXL/RDMA-based remote memory pools.
\update{To further demonstrate \systemname's versatility, we generalize its design from traditional containers for microVM-based agent workloads and introduce new optimizations, including browser sharing and a page cache bypassing mechanism.}
Our evaluation shows that \systemname achieves up to 7$\times$ reduction in P99 latency and 48\% memory savings for container-based functions. \update{When applied to LLM agents, it reduces the P99 latency by up to 58\% and memory usage by 61\% compared to state-of-the-art systems like E2B.}
\end{abstract}

\ccsdesc[500]{Computer systems organization~Cloud computing}
\ccsdesc[500]{Software and its engineering~Memory management}

\keywords{Serverless, Agent, CXL, Cold Start}


\maketitle

\input{content/intro}
\input{content/background}
\input{content/agent-back}
\input{content/overview}
\input{content/design}
\input{content/agent}
\input{content/implementation}
\input{content/discussion}
\input{content/eval}
\input{content/related-work}
\input{content/conclusion}

\begin{acks}
This work is supported by National Key Research \& Development Program of China (2022YFB4502004), Natural Science Foundation of China (62141216) and Tsinghua University Initiative Scientific Research Program, Young Elite Scientists Sponsorship Program by CAST (2022QNRC001).
\end{acks}

\bibliographystyle{ACM-Reference-Format}
\bibliography{base}







\end{document}

%% file: content/intro.tex
\section{Introduction}

Serverless computing, renowned for its fine-grained resource allocation and billing, enables applications to scale efficiently while optimizing resource utilization.
Given its transformative potential, serverless computing is now available across all major cloud service providers~\cite{ali-fc,amazone-lambda,google-cloud-run} and is applied across a broad range of sectors\cite{amazon-case-study,ao_sprocket_2018,carreira_cirrus_2019,jonas_occupy_2017,pu_shuffling_nodate}
\update{, such as the emerging large language model~(LLM) agents~\cite{deepresearch,mgx2025} workloads.}

However, despite the utopian vision of serverless computing, its real-world implementations face many practical limitations. 
Although serverless models suggest that functions could be transparently scheduled across data centers to maximize resource utilization, 
this idealized elasticity is often compromised by the functions' need for efficient access to local and dedicated background environments. 
This {\bf contradiction between computational elasticity and environment localization} poses challenges 
to the full realization of serverless's potential\update{, and becomes especially critical for complex applications like LLM agents that require high-density deployment and resource overcommitment.}

Precisely, the execution of a serverless function typically unfolds in three phases.
{\em (1)} The sandbox creation phase establishes an isolated sandbox (e.g., a container);
{\em (2)} The bootstrapping phase initializes the function, which may involve steps like launching Python/Java virtual machines;
and finally, {\em (3)} The execution phase executes instance-specific logic according to the inputs of this invocation.
The first two phases are essential for setting up an execution environment for each function invocation. 
However, they do not directly contribute to processing user inputs, thus their overheads (in both CPU time and memory consumption) should be minimized.

Unfortunately, studies show that the time required to create such dedicated environments can significantly exceed the actual execution time of a function, leading to the notorious ``cold start'' problem in serverless computing~\cite{silva_prebaking_2020,du_catalyzer_2020}. 
This issue has prompted the development of various caching techniques aimed at keeping functions warm and ready for immediate reuse~\cite{oakes_sock_nodate,akkus_sand_nodate,azure-warm-instance,aws-provisioned-concurrency}. 
However, these mechanisms necessitate the reservation of local resources, particularly memory, introducing a trade-off between initialization speed and resource costs, thereby diminishing the system's overall elasticity. 
For instance, commercial solutions like AWS Lambda's Provisioned Concurrency~\cite{aws-provisioned-concurrency}, which provide caching, deviate from the ideal pay-as-you-go model of cloud computing by charging for these reserved resources.

The requirement for execution environments also gives rise to other issues, such as ``memory stranding''~\cite{lagar_software_2019} and ``state duplication''~\cite{saxena_memory_2022}.  
Memory stranding occurs when a server's computational resources are fully engaged, but its memory remains underutilized.
State duplication arises when concurrent serverless functions require but cannot efficiently share identical states due to existing isolation mechanisms.
\update{
This is exacerbated when multiple instances utilize common toolkits or language runtimes, such as in the LLM agent workloads. 
Furthermore, the duplication is amplified in virtual machine (VM) environments where guest OS images and files are redundantly loaded.
}
Recent studies have indicated that these challenges lead to significant inefficiencies, with up to 50\% of memory resources being underutilized in the cloud~\cite{javadi_scavenger_2019,fuerst_memory-harvesting_2022} and an 80\% occurrence of state duplication~\cite{saxena_memory_2022}. 
These issues collectively pose significant obstacles to achieving genuine elasticity in serverless computing.

\subsection{Our Contribution}\label{sec:our-contribution}

\begin{table}[h]
\caption{The core components in current containers. \update{The evaluation setup is detailed in \S\ref{sec:eval-methodology}.}}
\label{tab:container-components}
\centering
\small
\begin{tabular}{@{}l|llcl@{}}
\toprule
                      & Unit    & Description                                                                                                                                                       & Overhead            & \systemname's Solution                                                                                                      \\ \midrule
\multirow{10}{*}{\rotatebox[origin=c]{90}{Sandbox}}  & Network & \begin{tabular}[c]{@{}l@{}}Isolated network environment, such as\\ independent ports, including a network\\ namespace and a virtual ethernet device.\end{tabular} & 80 ms $\sim$ 10 s   & \begin{tabular}[c]{@{}l@{}}Direct reuse as it does not\\ leak any data produced\\ during processing (\S\ref{sec:reused-kernel-objects}).\end{tabular}      \\ \cmidrule(l){2-5} 
                      & Rootfs  & \begin{tabular}[c]{@{}l@{}}Root filesystem for containers, including a\\ mount namespace and necessary \\ filesystems such as sysfs under /sys.\end{tabular}       & 10 $\sim$ 800 ms    & \multirow{4}{*}{\begin{tabular}[c]{@{}l@{}}Reuse with reconfigurations,\\ at a lower cost than creating \\
                      (\S\ref{sec:isolated-env-find-tune}).\end{tabular}} \\ \cmidrule(lr){2-4}
                      & Cgroup  & \begin{tabular}[c]{@{}l@{}}Resource isolation, such as CPU and\\ memory usage control.\end{tabular}                                                               & 30 $\sim$ 400 ms    &                                                                                                                       \\ \cmidrule(l){2-5} 
                      & Other   & \begin{tabular}[c]{@{}l@{}}Other isolated resources, such as time and \\ pid namespaces.\end{tabular}                                                              & \textless 1 ms      & Create with low overhead.                                                                                             \\ \midrule
\multirow{4}{*}{\rotatebox[origin=c]{90}{Process}} & Memory  & \begin{tabular}[c]{@{}l@{}}Post-initialized state of functions,\\ including loaded language runtime, \\ imported libraries and user code.\end{tabular}            & \textgreater 300 ms & \begin{tabular}[c]{@{}l@{}}Introduce new kernel\\ interfaces, bypass costly\\ memory copy (\S\ref{sec:mm-template-design}).\end{tabular}               \\ \cmidrule(l){2-5} 
                      & Other   & \begin{tabular}[c]{@{}l@{}}Multi-thread context, registers, sockets,\\ open file descriptors, etc.\end{tabular}                                                            & 3 $\sim$ 15 ms      & \begin{tabular}[c]{@{}l@{}}Handled by CRIU with\\ strong generality.\end{tabular}                                     \\ \bottomrule
\end{tabular}
\end{table}

In this paper, we present \systemname, a serverless computing platform crafted to minimize CPU time and memory costs associated with serverless execution environments.
The main idea is to share components of these environments across different functions and nodes as much as possible.
Rather than discarding this execution environment upon the completion of a function instance, \systemname cleanses and reallocates it to a repurposable sandbox pool.
Subsequent functions can reuse these sandboxes by attaching function-specific memory states, which are offloaded to a shared CXL or RDMA-based memory pool. 
This process, termed ``repurposing'' in \systemname, is depicted in Figure \ref{fig:high-level}.
Through innovative enhancements at the operating system (OS) and core libraries like Checkpoint/Restore In Userspace (CRIU)~\cite{criu},
this procedure typically takes less than 10 ms for containers.
Only updated (writable) memory pages are instantiated per invocation, while read-only pages are transparently shared across multiple instances and nodes.

\systemname's approach of container repurposing, while bearing similarities to existing caching solutions, distinguishes itself in two significant ways. 
Firstly, \systemname enables the transparent reuse of sandboxes not only across invocations of the same function, but also across different function types, regardless of whether the function runs in a container or microVM environment.
\systemname performs a systematic decomposition of sandbox creation overhead to determine which components are safely reusable and which require lightweight reconfiguration.
To achieve this, we have identified crucial kernel objects that contribute to the overhead of creating isolated sandboxes. 
Some of these components are directly reusable across different functions, while others necessitate a tailored reconfiguration that upholds cost-effectiveness without
compromising isolation properties relative to previous container-based systems.
These components are listed in Table \ref{tab:container-components} and will be discussed in detail in \S\ref{sec:isolation-challenge} and \S\ref{sec:isolated-env-find-tune}.

Secondly, \systemname offloads function-specific memory states to a shared memory pool. 
This strategy shares the states across multiple server nodes, effectively amortizing the overall expense. 
\systemname also supports an effective deduplication process that eliminates redundant states across different functions, potentially reducing memory usage by up to 48\% in our evaluated functions~(Table \ref{tab:evaluated-funcs} in \S\ref{sec:eval-methodology}).
However, operating system modifications are required to rapidly instantiate serverless functions from snapshots stored in remote memory pools.
This process resembles performing a ``fork'' but includes capabilities beyond those offered by current operating systems.
To support this, we have extended the kernel with an \textit{mm-template API}, enabling the creation of custom memory templates for each serverless function. 
These templates consist of reserved page tables that map to remote, potentially overlapping, non-contiguous memory segments within the pool, representing deduplicated snapshots of serverless functions.

\begin{figure}[ht]
    \centering
    \includegraphics[width=.4\linewidth]{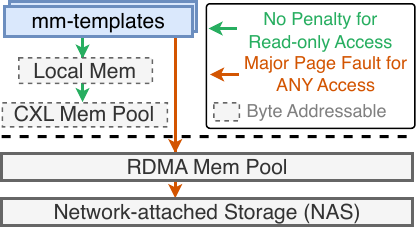}
    \caption{Multi-layer architecture of mm-template.}
    \label{fig:two-type-mem-pool}
\end{figure}

The flexibility of our design is one of its key strengths. As demonstrated in Figure \ref{fig:two-type-mem-pool}, \systemname integrates mm-templates with both CXL and RDMA memory pools, each offering distinct advantages. 
Notably, to leverage the byte-addressable capabilities of CXL, mm-templates enable direct reads for read-only pages, which introduces zero additional software-level overhead during execution.
Given our analysis showing that 24\% to 90\% of memory accesses are to read-only pages in serverless functions, this technique substantially accelerates processing times by avoiding unnecessary data copying or page faults, thereby significantly reducing latency.
Despite this, existing OS primarily utilizes CXL for memory expansion as per the capabilities of CXL 1.1, thus lacking support for memory sharing and deduplication required in multi-node environments introduced since CXL 2.0. 
To overcome this limitation, we have enhanced the OS to allow precise control over the mapping between virtual and physical addresses in CXL memory and supported both file-backed and anonymous memory mappings to maintain a process's complete state, including heap and stack areas.

Different from CXL, integrating RDMA-based remote memory follows a lazy paging strategy, where any access to remote memory triggers a major page fault that fetches a 4KB block online. 
Since all states in the memory pool are read-only, with write operations managed through copy-on-write mechanisms, a multi-layered architecture integrates seamlessly with our approach. 
This allows for the strategic placement of hot pages in the upper layers, such as CXL or even local memory, and cold pages in the lower layers, such as RDMA or even network-attached storage. 
The specific number of layers, as well as the cache eviction 
or page promotion strategies, are orthogonal to our core implementation.

Moreover, the ease of use is the most important attraction of serverless computing. To avoid the need for users to modify their code or recompile their applications, \systemname decides to \emph{transparently} incorporate repurposable sandboxes and mm-templates into the serverless platform.
It implements a new ``repurpose'' command in CRIU, based on the existing ``restore'' command.
This facilitates the seamless integration with many existing serverless platforms due to CRIU's widespread adoption in mainstream container runtimes like Docker\cite{docker} and Podman\cite{podman}.

\update{
While above optimizations primarily target container-based platforms, VM-based isolation remains the mainstream choice for public cloud providers due to its stronger security guarantees.
To demonstrate \systemname's versatility in VM-based environments, we apply and evaluate it on emerging LLM agent workloads, a challenging scenario characterized by dynamic invocation patterns and fluctuating resource demands.
This evaluation further led to two additional optimizations specific to the VM-based sandbox and LLM agent: page cache bypassing and browser sharing, which enhance startup efficiency and memory utilization.
Together, these extensions enable high-density, high-performance execution of LLM agent applications atop VM-based serverless platforms.

This paper makes the following key contributions:
\begin{itemize}
\item We analyze the performance bottlenecks and characteristics of agent and serverless applications on both container-~(\S\ref{sec:background-serverless-computing},\S\ref{sec:isolation-challenge},\S\ref{sec:mm-challenge}) and VM-based platforms~(\S\ref{sec:background-microvm},\S\ref{sec:case-study-agent},\S\ref{sec:agent-vm-challenge}).

\item We propose a repurposable sandbox design~(\S\ref{sec:isolated-env-find-tune}) that enables transparent sharing across heterogeneous function types, reducing per-invocation overhead.

\item We design and implement the mm-template API~(\S\ref{sec:mm-template-design}), which facilitates remote memory state sharing with minimal restoration latency and reduced memory duplication.

\item We introduce two optimizations for agent applications on VM-based platforms: (1) extending our sharing mechanism to browser processes to reduce execution latency~(\S\ref{sec:browser-share}), and (2) integrating virtio-pmem with mm-template to enable memory state sharing while mitigating page cache duplication~(\S\ref{sec:page-cache-mitigate}).

\item We develop \systemname, a serverless platform built atop faasd~\cite{faasd} and Cloud Hypervisor~\cite{cloud-hypervisor}, supporting both container and VM runtimes.

\item Our evaluation on container-based workloads demonstrates that:
(1) For prior serverless workloads, \systemname achieves up to $7\times$ and $16\times$ speedup in P99 end-to-end latency under concurrent loads compared with REAP~\cite{ustiugov_benchmarking_2021} and FaaSnap~\cite{ao_faasnap_2022}, two state-of-the-art (SOTA) lazy restoration methods, respectively; \systemname reduces peak memory usage by 48\% on average;
(2) A detailed comparison between real-world CXL and RDMA memory pools reveals that CXL provides more stable performance, particularly at P99, and offers up to a $3.51 \times$ speedup on execution time over RDMA.

\item Our evaluation on VM-based agent workloads demonstrates \systemname achieves at most 61\% memory savings and 58\% P99 latency reduction compared with the SOTA system, E2B, after adopting our optimizations, while keeping the same efficient startup latency.

\item We have open-sourced \systemname, which is available at \url{https://github.com/kvcache-ai/TrEnv-X}.
\end{itemize}
}

%% file: content/background.tex
\section{Background and Motivation}

\subsection{Dissaggregated Memory}
\textbf{CXL} Compute Express Link (CXL)~\cite{cxl-spec,das_sharma_introduction_2024} is emerging as a promising technology in the realm of high-speed server interconnects, notable for its low latency~\cite{zhang_partial_2023}, high bandwidth~\cite{sun_demystifying_2023,tang_exploring_2024}, and ability to enable shared memory access between servers.
Its adoption by industry giants~\cite{li_pond_2023,maruf_tpp_2023} also underscores its increasing significance and application in modern computing infrastructure.

Our research primarily concentrates on CXL's capability of sharing memory across different machines, which has been supported by multi-headed devices (MHD) even with CXL 2.0~\cite{jain_memory_2024}. Real-world support for this capability has been demonstrated in previous research~\cite{zhang_partial_2023}.
Collaborations between many companies~\cite{samsung,xconn,memverge,h3} have also led to a commercial solution that enabled up to 7.5 TB CXL-attached memory pool shared by up to 12 servers~\cite{cxl-demo}.
We only need read-only sharing capabilities, so a Type-3 device with multiple CXL 2.0 interfaces (MHD) is sufficient without the hardware coherency introduced in CXL 3.0.

CXL's fine-grained and efficient access capability means only the needed cache lines are retrieved, not entire pages. Thus, it provides a potentially faster alternative to RDMA.

\noindent\textbf{RDMA}
Prior to CXL, RDMA was the standard protocol for building disaggregated memory pools, which can be grouped into two main categories:
Firstly, solutions such as Infiniswap~\cite{gu_efficient_nodate} and Fastswap~\cite{amaro_can_2020} enable efficient swapping over RDMA.
Secondly, frameworks and runtimes like AsymNVM~\cite{ma_asymnvm_2020} and FaRM~\cite{farm} offer coarse-grained abstractions that resemble key-value stores or file interfaces.
\systemname tries to implement transparent and extensible in-kernel functionality to support various remote memory, including but not limited to RDMA.

\subsection{Serverless Computing}\label{sec:background-serverless-computing}
The very idea of serverless is fragmenting applications into smaller, independent tasks that can be dispatched to whichever node has available CPU resources, thereby enhancing resource utilization. However, the pursuit of this optimal elasticity often encounters two predominant barriers:
\begin{enumerate}
    \item A target node may sometimes lack sufficient memory, preventing task dispatch, due to the tight coupling between computational and memory resources.
    \item Additional overheads introduced by this fragmentation, such as cold starts and memory duplication.
\end{enumerate}
The adoption of disaggregated memory has emerged as a viable solution to the first issue, with studies like Fastswap~\cite{amaro_can_2020} and Pond~\cite{li_pond_2023} exploring its benefits.
Our research pivots towards harnessing CXL/RDMA-based shared memory pool to tackle the second barrier.

\begin{figure}[h]
    \centering
    \includegraphics[width=.65\linewidth]{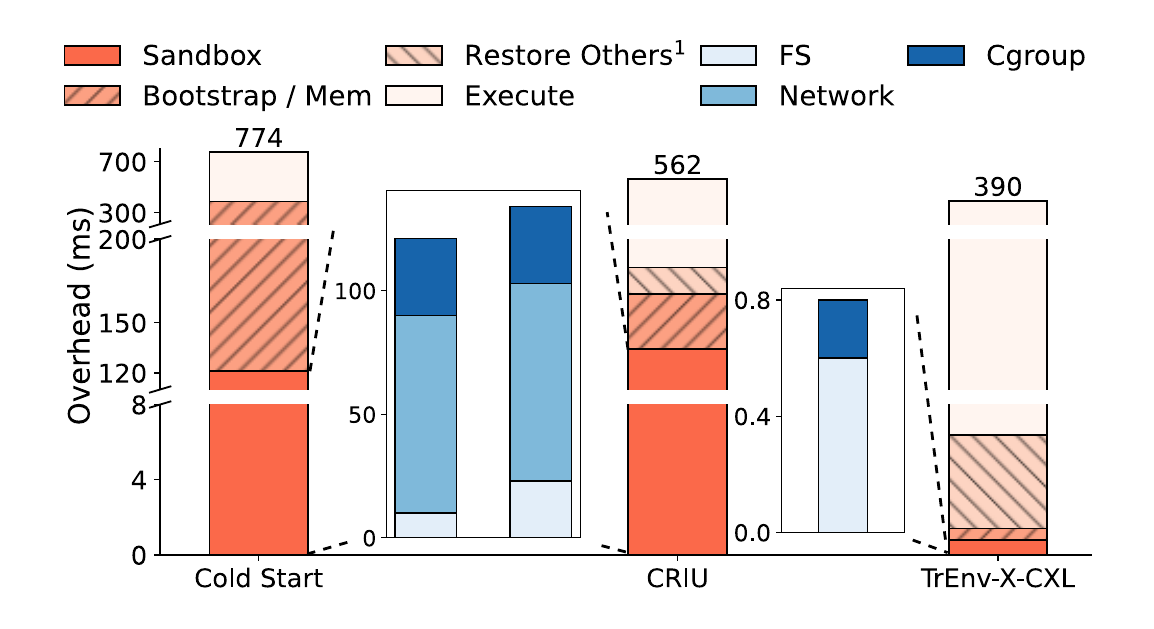}
    \begin{minipage}{\linewidth}
    \footnotesize{\update{$^1$ Restores other states of processes, such as recovering multi-thread context, reopening files, and re-establishing sockets.}}
    \end{minipage}
    \caption{Breakdown of the latency for a Python-based function, highlighting the overhead associated with the sandbox. \update{The evaluation setup is detailed in \S\ref{sec:eval-methodology}.}}
    \label{fig:serverless-breakdown}
\end{figure}

Referring to the ``Cold Start'' bar in Figure~\ref{fig:serverless-breakdown}, we identify two significant costs:
\begin{enumerate}
    \item The preparation of an isolated sandbox, including tasks like creating the rootfs.
    \item The cost involved in bootstrapping the function’s processes, such as launching interpreters and importing libraries. 
\end{enumerate}
These ancillary overheads overshadow the proportion of the function's actual productive execution time, given that many studies report that most serverless functions execute in less than one second~\cite{shahrad_serverless_nodate,joosen_how_2023}.
\update{Besides, the same cold start problem exists in LLM agent appliactions, especially for ones with short end-to-end latency, e.g., Blackjack in Table \ref{tab:agent-chara}.}

\update{
\subsection{MicroVM in Serverless}\label{sec:background-microvm}
Modern serverless platforms generally fall into two categories:
\begin{enumerate}
    \item Container-based platforms (e.g., OpenFaaS~\cite{openfaas}), which employ Linux namespaces and cgroups for lightweight isolation.
    \item VM-based platforms (e.g., Cloud Hypervisor~\cite{cloud-hypervisor}), which provide hardware-enforced isolation through virtual machines.
\end{enumerate}

While container-based frameworks offer efficiency and low overhead, their shared-kernel design weakens security. Containers rely on software-based isolation; thus, any privilege escalation or kernel vulnerability (e.g., CVE-2017-5123~\cite{docker_cve}) can allow an attacker to escape the container and compromise the host.

In contrast, VM-based serverless platforms encapsulate each function within a dedicated guest kernel, protected by hardware mechanisms such as extended page tables (EPT)~\cite{intel_manual}.
Although such designs mitigate security concerns, they introduce new inefficiencies, including longer startup latency and higher memory overhead from hypervisors.
Recent efforts purpose the microVM~\cite{agache_firecracker_nodate,zijun_rund_2022}, mitigating these costs by employing minimized guest kernels and simplified device models. It can achieve performance comparable to containers.
However, our experiments reveal a persistent inefficiency: page cache duplication among microVMs.
In serverless environments, multiple instances of the same function often access identical binaries and data, yet each VM maintains its own virtual block device and guest-side page cache.
As a result, identical files are redundantly cached across guest kernels and even duplicated within their corresponding image files on the host.

Despite their architectural differences, containers and microVMs share a fundamental similarity from the host’s perspective: both execute as host processes governed by Linux cgroups and namespaces.
Consequently, both suffer comparable overheads in preparing and managing isolated sandboxes.
}

\subsection{Existing Research}\label{sec:background-existing-research}
\textbf{Cold start overhead.}
Historically, caching mechanisms have been extensively studied and deployed to mitigate cold start overheads~\cite{fuerst_faascache_2021,roy_icebreaker_2022,shahrad_serverless_nodate}.
However, as discussed earlier, they introduce a trade-off between expediting initialization and reserving resources, which in turn impedes system elasticity.

Prior research efforts have thus been directed towards reducing the cost of caching by eschewing the reservation of active containers. 
Instead, they created snapshots or templates of a function's complete state immediately after initialization and stored them as files~\cite{silva_prebaking_2020,gvisor-cr,fc-cr}.
When a new instance needed to be started, it was restored from the snapshot, bypassing the bootstrapping phase.

Nevertheless, these strategies fail to fully conquer the aforementioned challenges. As highlighted by the CRIU bar in Figure~\ref{fig:serverless-breakdown}:
\begin{enumerate}
    \item The need to establish a new isolated sandbox still exists, thereby sustaining the associated overheads. Our findings also indicate that isolation costs escalate with concurrent cold starts. For example, concurrently initiating 15 instances results in a network setup time of 400 ms, aligning with prior studies~\cite{agile_mohan,oakes_sock_nodate}.
    \item the memory restoration overhead (``Mem'' in Figure \ref{fig:serverless-breakdown}) is still non-negligible due to costly data copying.
    In Figure~\ref{fig:serverless-breakdown}, we store the snapshot files in a local DRAM-based tmpfs to avoid the overhead of disk or network I/O.
    However, as we can see from the figure, memory copying alone takes over 60 ms during container bootstrapping, even for its small 60MB memory image. This overhead scales with the size of the memory image.
    For instance, a 360MB memory image would take over 220 ms to restore.
\end{enumerate}

Further optimizations have been suggested to alleviate
these challenges. But, according to our investigation, a holistic solution remains elusive.
Lightweight container technologies~\cite{oakes_sock_nodate}, for instance, mitigate the overhead of sandbox setup but do so at the cost of compromised isolation, making them unsuitable in multi-tenant contexts.
In practice, serverless platforms, such as OpenWhisk~\cite{openwhisk}, still resort to standard containers for a better level of isolation, or even opt for secure containers (i.e., VM)~\cite{gvisor-cr,agache_firecracker_nodate,zijun_rund_2022}, which could entail higher initialization costs.

Additionally, several ``lazy restore'' techniques have been proposed to hide the latency of data copying, such as recording and prefetch-based replay~\cite{ao_faasnap_2022,ustiugov_benchmarking_2021} or restoring state through efficient RDMA reads in a manner akin to a remote on-demand ``fork''~\cite{wei_no_2022}.
Nonetheless, these methodologies (1) merely defer the restoration overhead to the execution phase rather than reduce it, (2) do not leverage the potential of various remote memory pools, (3) focus on memory restoration but overlook the overhead associated with isolated sandboxes, such as namespaces and cgroups.

\newc{
\noindent\textbf{Duplicated page cache.}
Prior work has addressed similar issues by implementing exclusive caching between the storage cache (e.g., RAM on disk arrays) and the OS page cache~\cite{jones_geiger,bairavasundaram-x-ray}.
These solutions estimate the contents of the guest OS page cache based on file metadata or access patterns, and use this information to guide which blocks should be cached in a secondary storage layer. However, such gray-box approaches require the hypervisor to infer guest page cache behavior from indirect signals (e.g., VM exits), which often leads to inaccurate decisions.
Alternatively, content-based deduplication has been proposed to mitigate cache redundancy~\cite{sharma_singleton_2012}. This method avoids inference by directly scanning memory for duplicate content. However, it incurs significant CPU overhead and introduces latency due to the time required for scanning and matching duplicated pages.

More recently, Rund~\cite{zijun_rund_2022} adopts a para-virtualization method using virtiofs~\cite{virtiofs2025} with DAX (Direct Access) support, which maps the host page cache directly into the guest’s physical address space, \update{bypassing the guest page cache}.
While this design avoids duplication, it suffers from several limitations:
(1) The current virtiofs architecture relies on a separate daemon that serves guest file requests via a FUSE interface. This design mandates the use of shared memory (e.g., \texttt{memfd}) to back guest memory, which conflicts with copy-on-write mechanisms that are essential for offloading memory state across distributed remote memory pools.
(2) The DAX feature in virtiofs is not yet stable in official implementations. For example, virtiofsd~\cite{virtiofsd2025} removed DAX support in 2024, and Cloud Hypervisor deprecated its DAX-based virtiofs support as early as 2022.
}

%% file: content/agent-back.tex
\newc{
\section{Case Study: LLM Agent}\label{sec:case-study-agent}
\update{
The challenges of startup overhead and memory redundancy are particularly acute in VM-based serverless platforms. To evaluate the effectiveness and versatility of \systemname's design in this demanding context, we present a case study on LLM agents, a class of emerging workloads that exemplifies the dynamic and resource-intensive nature of future serverless applications.
}
\subsection{Basics of LLM Agents.}
Large language models (LLMs) have demonstrated substantial capabilities across a range of tasks, including natural language understanding and content generation~\cite{deepseekai2025deepseekv3technicalreport,openai2024gpt4technicalreport}.
These advances have catalyzed the emergence of a new class of LLM-driven applications for autonomous task completion, commonly referred to as agents (e.g., \cite{langchain2025,owl2025,hong2024metagpt}).
An agent is a system or program that leverages an LLM to solve user-defined tasks, incorporating capabilities such as reasoning, planning, and external tool usage.

\begin{table}[ht]
\caption{The characteristic of representative agents evaluated on Firecracker.}
\label{tab:agent-chara}
\small
\begin{tabular}{@{}l|llccc@{}}
\toprule
Agent              & Framework   & Description                             & E2E Lat$^*$ & Memory  & CPU Time \\ \midrule
Blackjack          & LangChain   & Play the Blackjack game.                & 3.2 s    & 74 MB   & 411 ms   \\
Bug fixer          & LangChain   & Fix the bugs in given code.             & 36.5 s   & 95 MB   & 809 ms   \\
Map reduce         & LangChain   & \update{Split and summarize a document.}           & 56.5 s   & 199 MB  & 1.2 s    \\
Shop assistant & Browser-Use & Select the ideal products on a website. & 140.7 s  & 1080 MB & 10.3 s   \\
Blog summary       & OWL         & Collect and summary blogs.              & 193.1 s  & 1246 MB & 56.8 s   \\
Game design        & OpenManus   & Implement a html-based game.            & 107.0 s  & 1389 MB & 7.5 s    \\ \bottomrule
\end{tabular}
\begin{flushleft}
    {\footnotesize $^*$ End-to-end latency.}
\end{flushleft}
\end{table}

\update{
To the best of our knowledge, this is the first work that aims to optimize serving platforms for LLM-based agents. Currently, no standard benchmark applications exist for evaluation.
We therefore selected representative agents from widely used open-source frameworks, including LangChain \cite{langchain2025}, Browser-Use~\cite{browser_use2024}, OWL~\cite{owl2025} and OpenManus~\cite{openmanus2025}.
The chosen agents encompass a broad spectrum of scenarios, covering diverse execution patterns (e.g., fixed, map-reduce, and ReAct), varied tool usage (e.g., browsers, RAG, and code interpreters), and different running characteristics (e.g., end-to-end latency and resource utilization).
}
We evaluate representative agents from widely-used open-source frameworks, including LangChain \cite{langchain2025}, Browser-Use~\cite{browser_use2024}, OWL~\cite{owl2025} and OpenManus~\cite{openmanus2025}.
LangChain offers the greatest flexibility for constructing agent applications, enabling users to compose custom workflows by integrating modular components. However, it typically requires manual programming to assemble an agent.
In contrast, Browser-Use integrates robust browser automation into the agent design. OWL and OpenManus provide out-of-the-box agents equipped with built-in tools, such as browsers and search engines.
\update{Table \ref{tab:agent-chara} summarizes the resource requirements of these agents deployed on a VM-based platform that leverages Firecracker's native snapshot mechanism.} Notably, we observe substantial variation in execution time and memory consumption across agents. Based on their resource profiles and structural complexity, these agents can be broadly categorized into two types:
\begin{enumerate}
    \item These agents use a minimal set of tools and exhibit low memory consumption and short end-to-end execution times. \update{For example, the ``Bug Fixer'' agent follows a simple static workflow: (a) it executes erroneous code using a Python interpreter, and (b) sends the output to an LLM API to obtain fix suggestions.}
    \item These agents often adopt the ReAct strategy~\cite{yao2023react} and are equipped with a diverse set of tools. For instance, the OWL agent employs a dynamic execution graph and integrates search engines, web browsers, and document parsers (e.g., for PDF files). Unlike lightweight agents, their execution is governed by real-time LLM decisions rather than a pre-defined static workflow.
\end{enumerate}

\begin{figure}[h]
    \begin{subfigure}{.28\linewidth}
        \includegraphics[width=\linewidth]{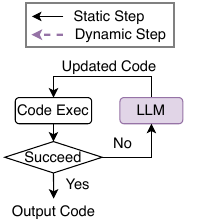}
        \caption{Bug fixer's static workflow.}
    \end{subfigure}
    \hfill
    \begin{subfigure}{.35\linewidth}
        \includegraphics[width=\linewidth]{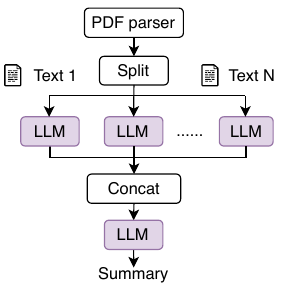}
        \caption{Map reduce workflow.}
    \end{subfigure}
    \hfill
    \begin{subfigure}{.35\linewidth}
        \includegraphics[width=\linewidth]{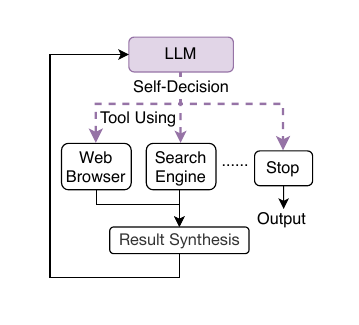}
        \caption{ReAct Agent with tool using.}
    \end{subfigure}
    \caption{Typical workflow of agent applications.}
    \label{fig:agent-application}
\end{figure}

\subsection{Serverless for Agents}
In the serverless computing model, applications are structured as individual \emph{functions}, each deployed independently and triggered by configurable events. The number of concurrently running function instances automatically scales in or out based on workload demands.
Furthermore, cloud providers typically charge only for the actual execution time of each function, often at sub-second granularity. This paradigm has gained widespread adoption, particularly for applications with dynamic workloads and variable resource requirements.

Agent-based applications naturally exhibit dynamic execution characteristics, making them well-suited for serverless deployment. First, their execution times are highly variable. The execution of an agent involves multiple steps that can be represented as a directed acyclic graph (DAG). As illustrated in Figure \ref{fig:agent-application}, agents can follow diverse execution paths, depending on factors such as the task definition, user input, tool availability, LLM-generated content, and LLM runtime performance. For instance, in our experiments, the end-to-end latency of the "MapReduce" agent ranges from 42 to 60 seconds. Despite using a fixed execution DAG and identical input across runs, the inherent non-determinism of LLM generation introduces significant variation in runtime.
Second, agent memory demands vary widely. Some agents utilize only lightweight tools, resulting in minimal memory usage. For example, the "MapReduce" agent only relies on a simple external tool to parse PDF documents. In contrast, other agents require more complex tools, such as web browsers, which lead to higher memory consumption. For instance, the "Blog Summary" agent launches a browser to capture snapshots of web pages when searching for blog content on specified topics.

Given these execution-time fluctuations and diverse resource requirements, serverless platforms offer a compelling infrastructure for hosting agent applications, leveraging their elastic scalability and fine-grained, usage-based billing.


\subsection{Costs of Serverless for Agents}\label{sec:agent-serverless-cost}
Our detailed analysis reveals that \update{\textbf{deploying agents on serverless platforms can incur substantial costs, sometimes comparable to the cost of LLM API calls}}. This finding is somewhat counterintuitive, as LLM invocation is typically considered the primary cost concern.
However, recent advances in LLM inference techniques, such as batching~\cite{yu_orca_nodate} and disaggregation~\cite{zhong_distserve_nodate}, have significantly reduced the cost of LLM usage. For instance, the price of GPT-4o was halved between 2024 and 2025.

LLM billing is based on tokens, which are also the fundamental processing unit for LLMs. Depending on the tokenizer, a token may represent a word, part of a word, or a phrase. Each LLM call incurs costs based on the number of input tokens (prompt tokens) and output tokens (completion tokens), which are typically priced separately.
Given an input token length $L_{in}$ and output token length $L_{out}$, the total cost of an LLM call is calculated as:
\begin{equation}
    C_{LLM}=L_{in}\times P_{in} + L_{out}\times P_{out}
\end{equation}
where $P_{in}$ and $P_{out}$ denote the prices per input and output token, respectively. Table \ref{tab:llm-token-usage} reports the token usage across various agents.

\begin{table}[ht]
\caption{The LLM token usage in representative agents.}
\label{tab:llm-token-usage}
\begin{tabular}{@{}l|cc@{}}
\toprule
Agent          & Input Tok & Output Tok \\ \midrule
Blackjack      & 1690       & 8          \\
Bug fixer      & 1557       & 530        \\
Map reduce     & 8640       & 2644       \\
Shop assistant & 43185      & 1494       \\
Blog summary   & 49398      & 2703       \\
Game design    & 75121      & 2098       \\ \bottomrule
\end{tabular}
\end{table}

Most existing public serverless platforms (e.g., AWS Lambda, Alibaba Function Compute and Google Cloud Run~\cite{amazone-lambda,ali-fc,google-cloud-run}) charge based on execution time and allocated resources.
For example, AWS Lambda bills at \$ $1.67\times 10^{-8}$ per millisecond per GB of memory.
For a single function invocation with end-to-end execution time $T$ and memory allocation $M$, the cost is computed as:
\begin{equation}
    C_{s}=T\times P_s\times M
\end{equation}
where $P_s$ refers to the unit price per second per GB.

\begin{figure}[ht]
    \centering
    \includegraphics[width=0.9\linewidth]{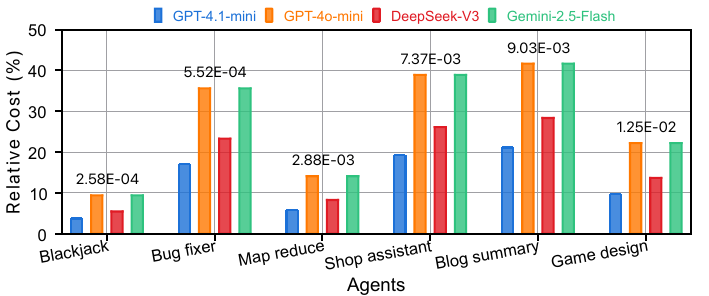}
    \caption{\update{The proportion of Serverless cost ($C_s$) relative to the total cost ($C_{LLM} + C_s$), where the upper value denotes the total absolute cost associated with the Gemini 2.5 Flash model.}}
    \label{fig:relative-price}
\end{figure}

Using this cost model, we analyze the total cost of running various agents (as summarized in Table \ref{tab:agent-chara}), and present the results in Figure \ref{fig:relative-price}. We observe two key findings:
\update{
Firstly, Serverless execution can represent a substantial portion of total expenditure. In our measurements, it accounted for as much as 42\% of overall costs for certain agents, which is comparable to that associated with LLM calls.}
Second, complex agents incur higher serverless costs. Their latency is typically split between (1) waiting for LLM outputs and (2) using external tools (e.g., web browsers). Unlike lightweight agents, which spend most of their time waiting on LLM responses, complex agents perform significant non-LLM operations, thereby accumulating longer active execution time and higher serverless charges.
As user tasks grow more complex, we expect the prevalence of complex agents to increase. Therefore, optimizing serverless cost efficiency, especially for agent-centric applications, is becoming increasingly important.

\subsection{Limitation of Serverless for Agents}\label{sec:serverless-agent-limitation}
Further analysis of agent workloads reveals two primary contributors to the high cost of serverless deployment.
First, CPU utilization is generally low. Although one of the key advantages of serverless computing is its pay-as-you-go pricing model, users are still billed for CPU resource allocation over the entire execution duration. However, agent workloads typically do not require exclusive CPU usage throughout their runtime. In fact, as shown in Table \ref{tab:agent-chara}, agents often utilize less than 25\% of allocated CPU resources during execution. This inefficiency is largely due to I/O-bound behavior, where agents spend a significant portion of their time waiting for external I/O operations rather than actively performing computation.

Second, memory consumption remains high for certain (especially complex) agents, even though the measurements reported in Table \ref{tab:agent-chara} already adopt snapshotting techniques. Specifically, memory that is (1) unused after initialization or (2) read-only is excluded from the usage statistics. Thus, the reported memory footprint reflects only the dynamic memory allocated during runtime.
A major contributor to this runtime memory usage is \textbf{page cache duplication between the guest and host operating systems.} For example, in the ``Blog Summary'' agent, approximately 500 MB is consumed by the guest kernel's page cache and another 500 MB by the host kernel's page cache. This duplication arises from the para-virtualized storage device used in Firecracker.
When a file is accessed by an application inside the guest, the data is cached in the guest page cache. The access is then emulated by the host hypervisor, causing the identical data to be cached again in the host page cache. This redundant caching during execution leads to inefficient memory utilization, as both the guest and host maintain separate copies of the same content.

}

%% file: content/overview.tex
\section{Overview of the Container-Based Platform}\label{sec:overview}

\begin{figure}[!t]
    \begin{minipage}[b]{.42\linewidth}
        \centering
        \includegraphics[width=\linewidth]{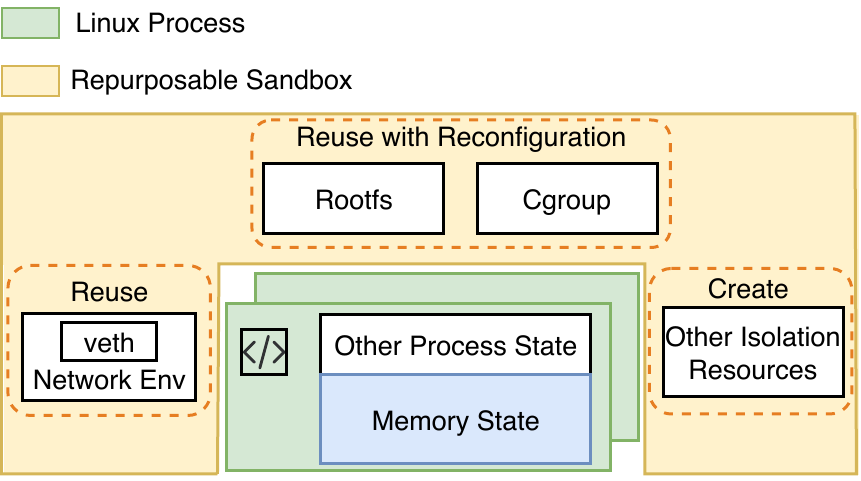}
        \caption{\systemname's view of a container. It consists of two parts:  a repurposable sandbox, which is comprised of isolation resources such as the root filesystem~(rootfs); and a group of processes,  the critical component of which is the memory state.}
        \label{fig:container}
    \end{minipage}\hfill
    \begin{minipage}[b]{.56\linewidth}
        \centering
        \includegraphics[width=\linewidth]{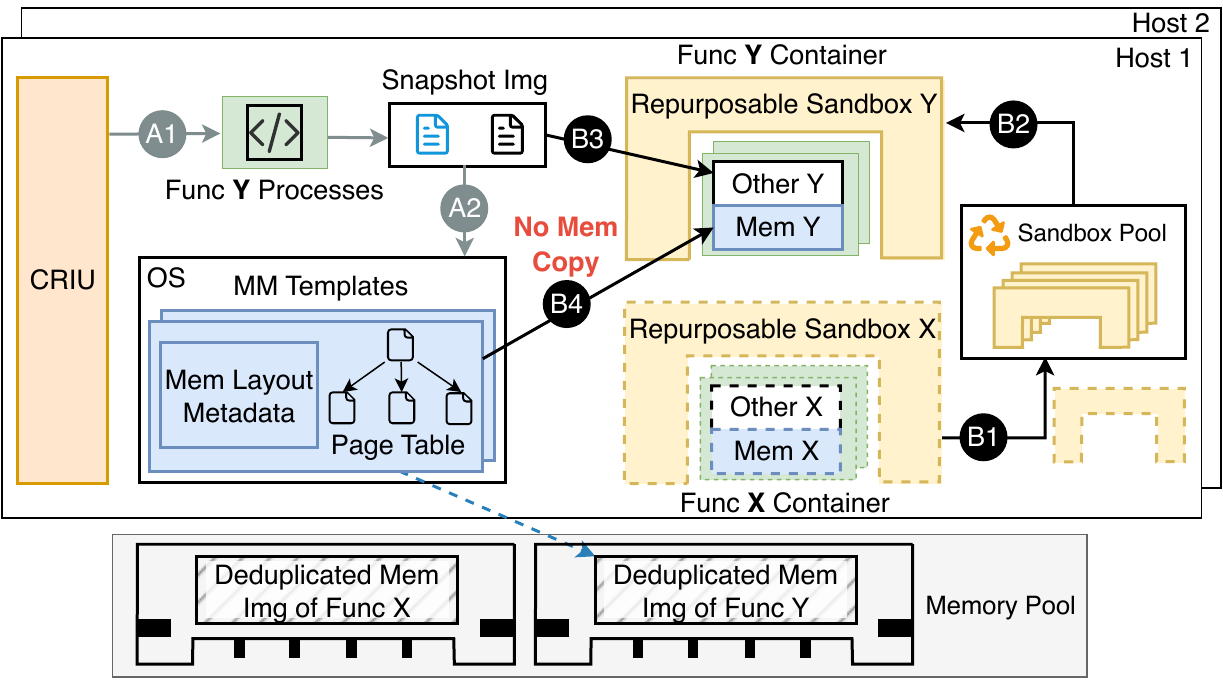}
        \caption{Overview of \systemname. Steps A1-A2 refer to offline preprocessing, while steps B1-B4 refer to online repurposing.}
        \label{fig:high-level}
    \end{minipage}
\end{figure}


The key idea of \systemname is an innovative repurposing strategy for reusing sandboxes and memory states of functions.
In a traditional caching strategy, the cached instance is limited to being reused by the same function.
\systemname diverges from this by enabling an efficient transition from an idle function instance to any one of the pending functions, regardless of its type. Thus, it achieves a universal sandbox pool. 
By moving away from the rigid, function-type-specific model to a more flexible, function-type-agnostic one, our design substantially enhances the elasticity of serverless computing.
Crucially, \systemname accomplishes this with the same or stronger level of security and isolation compared with previous systems.

\begin{figure}[ht]
\begin{minipage}[b]{.52\linewidth}
        \centering
        \includegraphics[width=\linewidth]{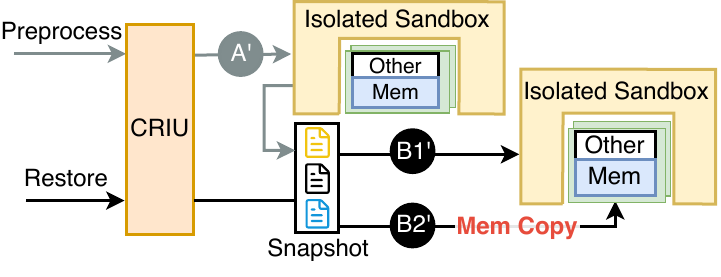}
        \caption{Two phases of CRIU in current container runtime. Step A$^{\prime}$ is the preprocessing phase, while step B1$^{\prime}$-B2$^{\prime}$ are the online restoration phase.}
        \label{fig:criu-two-phase}
\end{minipage}\hfill
\begin{minipage}[b]{.43\linewidth}
        \centering
        \includegraphics[width=\linewidth]{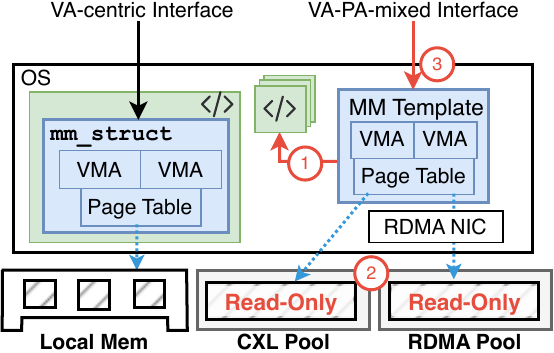}
        \caption{Three highlights in mm-template compared to the \texttt{mm\_struct} in Linux.}
        \label{fig:mm-template}
\end{minipage}
\end{figure}

Implementing such a transparent and efficient state transition necessitates OS kernel changes and updates to vital serverless infrastructures, such as CRIU.
To further elucidate our concept and its accompanying challenges, Figure \ref{fig:criu-two-phase} and Figure \ref{fig:high-level} present a detailed comparison of the function initialization workflows between current C/R (checkpoint and restore) systems and \systemname.

The current workflow involves two phases: a preprocessing phase and an online restoration phase.
As depicted in step A$^{\prime}$ of Figure \ref{fig:criu-two-phase}, the preprocessing phase involves creating snapshots for each function and storing them as files. These snapshots, representing the post-initialization state of a function, are used for subsequent restorations.
When a function is invoked, the system performs two steps to start a new container: (B1$^{\prime}$) it recreates a new sandbox (e.g., rootfs and cgroup) based on metadata in the snapshot image;
and then (B2$^{\prime}$) it restores the processes states. A significant challenge in step B2$^{\prime}$ is restoring the memory state, which requires numerous \texttt{mmap()} system calls to recover virtual memory layouts and costly data copying to reload memory contents.

While maintaining both phases, \systemname introduces improvements to streamline the restoration phase into the more efficient repurposing phase.
In the preprocessing phase,  \systemname generates snapshots for each function~(step A1 in Figure \ref{fig:high-level}) akin to the original CRIU. 
Yet, rather than saving memory states as files,
\systemname first deduplicates them into consolidated images stored on remote memory pools, then constructs one memory template (mm-template) \emph{per Linux process} in the function, based on these snapshots~(step A2).


As shown in Figure \ref{fig:mm-template}, each mm-template is an in-kernel object that has a structure similar to how traditional memory state is maintained by the kernel for each process~(i.e., \texttt{mm\_struct}).
It has three unique features.
\Circled{1} Not bound to a particular process but can be attached dynamically into any active process.
\Circled{2} Treat all remote memory read-only and enable copy-on-write.
\Circled{3} Fine-grained control over page tables to map virtual address~(VA) to physical address~(PA) on remote memory.
It only contains the metadata of the memory state, such as a page table and virtual memory layouts~(i.e., \texttt{vm\_area\_struct}). Hence, its size is small~(e.g., \textless{} 1 MB).
By only copying this metadata instead of large memory images during function restoration, the overhead is \update{reduced considerably}.
Additionally, copying from the same mm-template enables seamless memory sharing among the instances of the same function and even across different hosts.
\systemname leverages these mm-templates to expedite function transitions.
As shown in Figure \ref{fig:high-level}, the online restoration phase in \systemname consists of four steps.
\begin{enumerate}[label=(B\arabic*)]
    \item Upon completion of an instance of function X,  \systemname cleans its sandbox and puts it into a pool instead of discarding it. This step includes terminating existing processes within the container, ensuring any residual state from the predecessor will not be kept for security reasons.
    \item When a function Y's invocation is pending, \systemname selects a sandbox from the pool and repurposes it into the pending type Y.
    This includes applying a unique overlay filesystem of function Y~(more in \S\ref{sec:rootfs-reconfig}) and restoring the corresponding cgroup limits.
    Note that \systemname still offers all isolation components~(e.g., cgroup and namespaces) as with a standard Docker container to every function instance.
    \item \systemname issues a ``repurpose'' request to CRIU.
    CRIU helps the restored processes of function Y to join the repurposed sandbox and recover other process states except for the memory state, such as calling \texttt{clone()} to restore multi-thread context. This repurpose-and-join pattern allows efficient reuse of sandbox units like namespaces and cgroups.
    \item \systemname attaches the mm-template to the restored process of function Y, efficiently restoring its memory state.
    Note that \systemname executes the aforementioned steps \emph{transparently}, requiring no code modifications or recompilation from users.
\end{enumerate}
However, the implementation of the above steps each has its challenges.

\subsection{Challenge on Reusing Isolated Sandbox}\label{sec:isolation-challenge}
As shown by the breakdown analysis in Figure \ref{fig:serverless-breakdown}, the cost of creating a new sandbox predominantly consists of three parts, 
namely for the rootfs, network environment, and cgroup.
While network namespaces and virtual network devices can be safely reused without leaking private data from execution, reusing other resources presents unique challenges.

Each function comes with unique dependencies in its rootfs, such as specific libraries or files. 
This diversity, along with the requirement of isolation, complicates the possibility of a universal sandbox that can be seamlessly repurposed across functions.
Consequently, there is a need for a mechanism that allows rapid filesystem transitions to accommodate these varying dependencies while minimizing storage overhead.
Previous work~\cite{li_help_nodate} proposed to mount the dependencies of different functions into the sandbox in a read-only manner.
However, it is essential to ensure \emph{write access} to all paths in the rootfs, which is crucial for a practical serverless platform, as emphasized by Alibaba~\cite{zijun_rund_2022} and Amazon~\cite{brooker_-demand_nodate}.

Cgroups play a critical role for resource isolation, including limiting the usage CPU, memory and block device.
Its typical workflow involves three steps: (1) create a cgroup for each sandbox (2) spawn container or VMM processes, and (3) move the processes into the newly created cgroup. The third step is referred to as cgroup migration in Linux.
Prior studies focused on alleviating the bottlenecks in the first step, such as maintaining a cgroup pool to amortize the creation overhead~\cite{oakes_sock_nodate,zijun_rund_2022}.
Our investigations reveal that the cgroup migration typically incurs latency ranging from 10 to 50 ms, while the cgroup creation latency ranges from 16 ms to 32 ms.
Therefore, there is a significant need to address and minimize the overhead stemming from cgroup migration.

\subsection{Challenge on Implementing mm-template}\label{sec:mm-challenge}
The implementation of memory templates necessitates substantial modifications to existing OS memory management interfaces. 
In this section, we highlight three challenges associated with the implementation.
First, mm-template must be extensible and capable of {\bf transparently} supporting access interfaces of different types of remote memory.
For example, CPUs can directly access CXL memory via load/store instructions, RDMA employs a message queue model, and NAS uses a block-based I/O interface. 
The mm-template should accommodate these different interfaces, establishing an extensible framework that can utilize multi-layer memory pools. 

Second, the system should {\bf minimize} additional overhead. Traditional memory systems often adopt a lazy or on-demand approach, triggering memory loading during execution via page faults~\cite{amaro_can_2020} or a userspace daemon~\cite{ustiugov_benchmarking_2021} via userfaultfd.
These approaches are acceptable in previous implementations because the predominant component of their latency remains in I/O~(e.g., 60 \textmu s for SSD and 6 \textmu s for RDMA). 
However, the low-latency, byte-addressable features of CXL memory make it crucial to avoid additional overheads, such as page faults, which could drastically degrade performance and obscure its advantages.

Finally, the OS needs to be extended to {\bf support shared CXL memory}.
Current OS memory interfaces are designed primarily for single-host memory expansion (as for CXL 1.1) and hence fall short in supporting cross-host sharing and diverse memory mappings that \systemname requires.

Presently, two main approaches are used for incorporating CXL memory in Linux:
\begin{enumerate}[label={(\arabic*)}]
    \item Expose the CXL memory as a CPU-less NUMA node~\cite{sun_demystifying_2023,li_pond_2023}.
    \item Designate CXL memory as a special device file~(e.g., \texttt{/dev/dax0.1}) and pass it to \texttt{mmap()}, with the help of direct access~(DAX)~\cite{dax2025} drivers.
\end{enumerate}

\begin{figure}[ht]
    \begin{subfigure}{0.33\linewidth}
    \includegraphics[width=\linewidth]{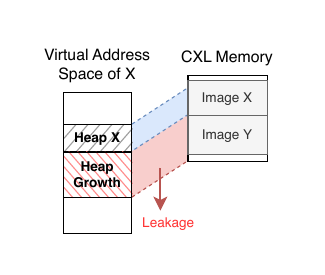}
    \caption{Vanilla DAX}
    \label{fig:dax-memory-leakage}
    \end{subfigure}
    \begin{subfigure}{0.33\linewidth}
        \includegraphics[width=\linewidth]{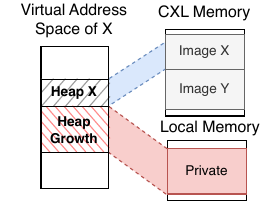}
    \caption{\systemname}
    \label{fig:our-memory-leakage}
    \end{subfigure}
    \caption{\newc{The result of heap growth during execution after restoring the heap area of function X on CXL memory.}}
    \label{fig:memory-leakage}
\end{figure}

The CPU-less NUMA approach is typically suitable only for memory expansion. Due to the Linux memory allocator's design for physical address transparency, it prevents the specification of physical addresses during allocation. Hence, it is challenging to coordinate multiple hosts to share CXL memory pages.
Although the DAX method permits setting physical offsets on CXL memory devices via \texttt{mmap}, it has its own set of limitations:
\begin{enumerate}[label={(\arabic*)}]
    \item The DAX driver does not support private mappings, which means it does not support copy-on-write for CXL memory pages.
    \item DAX is incompatible with regular-file-backed and some anonymous memory mappings, which are essential for critical process regions like heap and stack.
\end{enumerate}
\newc{As in Figure \ref{fig:dax-memory-leakage},} enforcing mm-template to restore heap areas through the DAX method could inadvertently cause heap growth~(e.g., \texttt{brk}) to jump into adjacent CXL memory ranges, posing risks of memory disclosure or data corruption.

%% file: content/design.tex
\section{Design of Container-based Platform}
In this section, we describe how \systemname solves the challenges described in \S\ref{sec:overview}.


\subsection{Memory Template Design}\label{sec:mm-template-design}
To enhance transparency, we have implemented the mm-template within the OS kernel and integrated it into CRIU. This integration allows serverless applications to leverage the advantages of memory pools without the need for code modifications or recompilation.
For extensibility, the mm-template supports various memory pool backends including CXL and RDMA.
During the preprocessing phase, CRIU creates and fills an mm-template for each process in the serverless function.
There is a built-in page table in the mm-template, which can be manipulated directly by its interface, i.e., \texttt{mmt\_setup\_pt}.
The semantics of this interface is to set up page table entries (PTEs) in the mm-template, to point to a particular memory region on remote memory.
The PTEs saves essential information including remote addresses and a special bit that identifies the type of memory pool.
For slower backends like RDMA, \systemname adopts a lazy approach, marking the corresponding PTEs as invalid. Thus, page faults are triggered during execution, allowing the kernel to identify these pages via the special bit in the PTE and route them to appropriate memory pool backends. These backends then allocate local pages and use the remote address in the PTE to load content from remote pools through their specific interfaces and implementations.

\begin{figure}[ht]
    \centering
    \includegraphics[width=.8\linewidth]{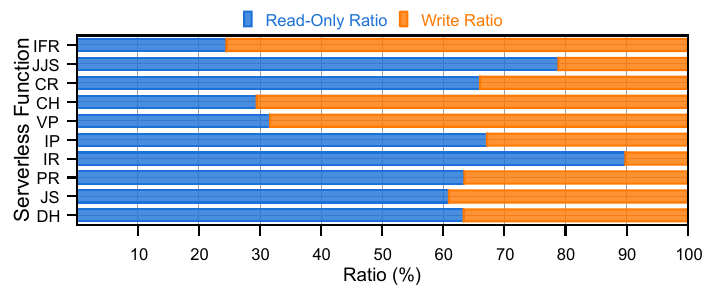}
    \caption{\newc{The read-only and write memory ratio of serverless functions in Table \ref{tab:evaluated-funcs}, using snapshotting.}}
    \label{fig:read-only-ratio}
\end{figure}

In contrast to remote memory pools like RDMA that are not directly accessible and incur page faults during execution, \systemname leverages low-latency, byte-addressable CXL memory to eliminate software-level overheads for read access.
During preprocessing, mm-template preconfigures valid page table entries (PTEs) that directly map to shared snapshot images in CXL memory. This enables the CPU to access data via standard load instructions, avoiding context switches and minor page faults. By shifting the overhead of on-demand paging to the preprocessing phase, \systemname achieves zero-cost read access during execution.
To enable memory sharing across function instances without duplicating the entire memory image, \systemname adopts a copy-on-write (CoW) strategy. When setting up PTEs for CXL-backed memory, write protection is applied during preprocessing. A write access at runtime triggers a page fault, at which point the page is copied to local RAM, preserving isolation across instances.
\newc{
To assess the potential for read-only sharing and the cost of CoW, we evaluate the functions listed in Table \ref{tab:evaluated-funcs}.
For each function, we take a snapshot after initialization, restore a single instance from the snapshot, and then execute a complete invocation. After execution, we count the number of memory pages that were read and written, respectively. The results, shown in Figure \ref{fig:read-only-ratio}, demonstrate that a significant portion of memory remains read-only, indicating that CoW can preserve efficiency while maintaining the integrity of single copy of remote memory images.
}

\begin{figure}[tb]
    \centering
    \frame{\includegraphics[width=.8\linewidth]{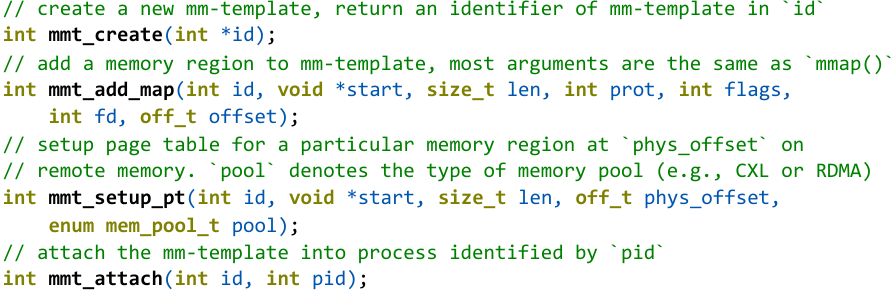}}
    \caption{Core API of mm-template.}
    \label{fig:mm-template-api}
\end{figure}

\systemname also introduces a custom kernel driver to remove the limitations of the standard DAX driver, which requires all CXL mappings to be associated with a device file (e.g., \texttt{/dev/dax0.1}). \newc{This requirement conflicts with anonymous mappings (such as heap memory), which are inherently file-independent, and prohibits mapping CXL memory to other files, such as the \texttt{.data} section of executables.
These restrictions stem from the design of the device-DAX driver, which configures PTEs at run time via callbacks functions associated with memory mappings during creation (i.e., \texttt{mmap}).
In the absence of a callback function association, the kernel defaults to allocating local memory pages, making transparent offloading to CXL infeasible.
The new driver in \systemname enables flexible memory management by allowing mm-template to explicitly set PTEs in advance.}
As the core mm-template API illustrated in Figure \ref{fig:mm-template-api}, the preprocessing phase uses \texttt{mmt\_add\_map} to create virtual memory areas, followed by \texttt{mmt\_setup\_pt} to (1) translate offsets to physical CXL addresses, (2) set valid write-protected PTEs, and (3) pin the corresponding pages.
This mechanism eliminates reliance on the DAX device and allows memory regions to be configured ahead of execution.
\newc{Thus, mm-template can support both anonymous and regular-file-backed mappings on CXL memory. As shown in Figure \ref{fig:our-memory-leakage}, heap areas reside in CXL memory after being attached via \texttt{mmt\_attach}, while subsequent heap expansion will default to local memory allocation, avoiding interference with other shared CXL regions.}

\begin{figure}[t!]
    \centering
    \includegraphics[width=.6\linewidth]{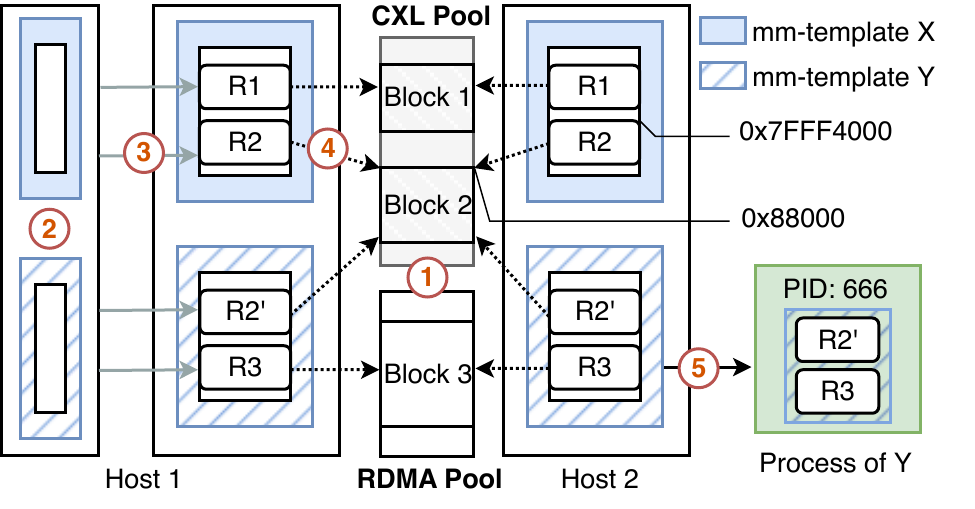}
    \caption{A simplified example of using mm-template.}
    \label{fig:mm-template-example}
\end{figure}

\noindent \textbf{Usage example of mm-template.}
Due to the transparency of mm-template, using CXL and RDMA only differs in the \`{}\texttt{pool}\`{} argument passed to \texttt{mmt\_setup\_pt} for PTE preconfiguration.
Most of the complexity is hidden in the kernel.
Figure \ref{fig:mm-template-example} shows the workflow to use mm-template, involving two single-process functions.
Initially, during the \emph{offline} preprocessing phase, the system generates a snapshot for each function using CRIU. \systemname then deduplicates these snapshots, resulting in a consolidated image stored in memory pools~(step \Circled{1} in Fig \ref{fig:mm-template-example}).
As shown in Fig \ref{fig:mm-template-example}, the snapshots for functions X and Y include a duplicated region (R2 and R2$^{\prime}$, respectively) mapped to the same Block 2 on remote memory. Additionally, \systemname logs the physical start offset of each memory block on the memory pool, such as 0x88000 for Block 2, which comprises, for example, 4 pages. 
This physical offset acts as a machine-independent pointer. 


Subsequently, each host constructs one mm-template for each process in the function~(step \Circled{2}). 
For example, \systemname calls \texttt{mmt\_create(\&X)} to initiate a memory template for function X. 
This template is then populated with specific virtual memory mappings using the \texttt{mmt\_add\_map} API~(step \Circled{3}).
For example, to allocate the private read-only anonymous mapping R2 in X's template, \systemname executes \texttt{mmt\_add\_map(X, 0x7FFF4000, 4*PAGE\_SIZE, PROT\_READ, MAP\_PRIVATE, -1, 0)}. 
The address 0x7FFF4000, derived from the snapshot image, represents the original virtual memory address of R2 in the checkpointed process of X.
The virtual memory mappings are then linked to the remote memory pool via \texttt{mmt\_setup\_pt} API~(step \Circled{4}).
For example, \texttt{mmt\_setup\_pt(X, 0x7FFF4000, 4*PAGE\_SIZE, 0x88000, CXL)} associates R2 in X's template with Block 2 on CXL memory.
Our new kernel driver helps to convert the offset 0x88000 to the corresponding physical address on CXL memory and install valid PTEs.
For the RDMA pool, such as R3 in mm-template Y, it installs invalid PTEs with the address in the RDMA memory pools, then loads memory pages in subsequent page faults.

During the repurposing phase~(i.e., critical path), \systemname attaches the mm-template to the process to be restored, via \texttt{mmt\_attach} API~(step \Circled{5}).
Each mm-template can be attached \emph{multiple times}, and it only copies the metadata, e.g., page tables, instead of the memory pages.
For instance, to restore the memory state of a process of function Y whose process ID~(PID) is 666, \systemname executes \texttt{mmt\_attach(Y, 666)}.

\subsection{Repurpose Isolated Sandbox}\label{sec:isolated-env-find-tune}
Two challenges related to isolated sandbox were mentioned in \S\ref{sec:isolation-challenge}. The first is that varying functions require different dependencies, making the reuse of rootfs non-trivial. The second is that the overhead of cgroup migration is non-negligible and cannot be solved by current approaches.

\subsubsection{Rootfs Reconfiguration}\label{sec:rootfs-reconfig}
\begin{figure}[t]
    \centering
    \includegraphics[width=.7\linewidth]{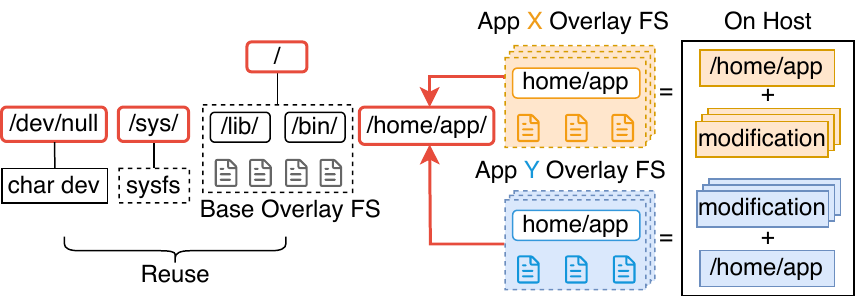}
    \caption{An example of rootfs reconfiguration. The red box indicates the mountpoint.}
    \label{fig:overlayfs}
    \vspace{-1ex}
\end{figure}

The standard container rootfs typically consists of multiple mountpoints within a per-container mount namespace~(mntns), often using a base union filesystem like overlayfs for the root directory.
Notably, many dependencies (e.g., glibc and language interpreters) are common across functions. Instead of switching the entire rootfs, our approach focuses on swapping only the function-specific dependencies.
By exploiting the flexibility of Linux mountpoints, we overmount another union filesystem atop an existing path, effectively ``replacing'' its contents, as shown in Figure \ref{fig:overlayfs}.
More precisely, there are three steps to reconfigure the rootfs between two different functions.
First, \systemname purges file modifications from previous instances. As overlayfs is a copy-on-write filesystem, all modified files are recorded in the upper directory. The purging process involves deleting all the files in the upper directory and remounting the overlayfs to flush stale inode cache.
Second, \systemname unmounts the function-specific overlayfs of previous function from the mntns.
Third, \systemname mounts the function-specific overlayfs of new function into the mntns.
\systemname further enhances this procedure by (1) executing the first step asynchronously, and (2) maintaining a pool of function-specific overlayfs, instead of discarding them after unmounting.

\noindent \underline{\emph{Compared with Cold Start}}. Typically, preparing a rootfs for containers from scratch requires many system calls, including more than 9 \texttt{mount}, 6 \texttt{mkdev}, 6 \texttt{mknod} and 1 \texttt{pivot\_root} calls to Linux.
For example, it involves mounting a sysfs under \texttt{/sys} and making a character device under \texttt{/dev/null}.
\systemname, in contrast, requires only 2 \texttt{mount} at minimum for function-specific dependencies and \texttt{/proc}, efficiently reusing other mountpoints. 

\noindent \underline{\emph{Compared with Lightweight Container}}. Many prior studies have adopted lightweight containers to overcome the isolation bottleneck.
For example, SOCK\cite{oakes_sock_nodate} has proposed a \emph{lean container}, using \texttt{chroot} and read-only bind mounts to set up its rootfs rapidly. However, according to the Linux manual page \cite{linux-chroot-man}, \texttt{chroot} is not intended for full process sandboxing, and
Sun \emph{et al.}~\cite{yuqiong_security_2018} have underscored its security limitations, which is susceptible to several potential ``escape'' attacks.
In contrast, \systemname still adopts the mount namespace as standard containers, which provides more robust isolation. 

\noindent \underline{\emph{Compared with Zygote Container}}. Prior research, such as Pagurus~\cite{li_help_nodate}, suggests a ``zygote'' container that can help various functions.
In Pagurus, each function-specific directory is assigned a unique owner, and the user ID of the function process is dynamically adjusted to access the corresponding directory.
However, it merely provides \emph{read-only} access, a limitation also found in lightweight containers.
\systemname, conversely, leverages the copy-on-write feature of overlayfs to emulate an ideal single-tenant environment with no restrictions on writable paths.
While zygote containers improve the hit ratio of idle containers compared with conventional caching strategies, one zygote can only help certain types of functions with similar dependencies and the same language.
\systemname, on the other hand, supports seamless transitions to arbitrary functions, even in different languages.
Moreover, zygote containers require additional initialization steps when helping other functions, such as loading unique libraries and user codes, leading to longer latency than \systemname.

\subsubsection{New Cgroup Feature Utilization}
\begin{figure}[htb]
    \centering
    \includegraphics[width=.7\linewidth]{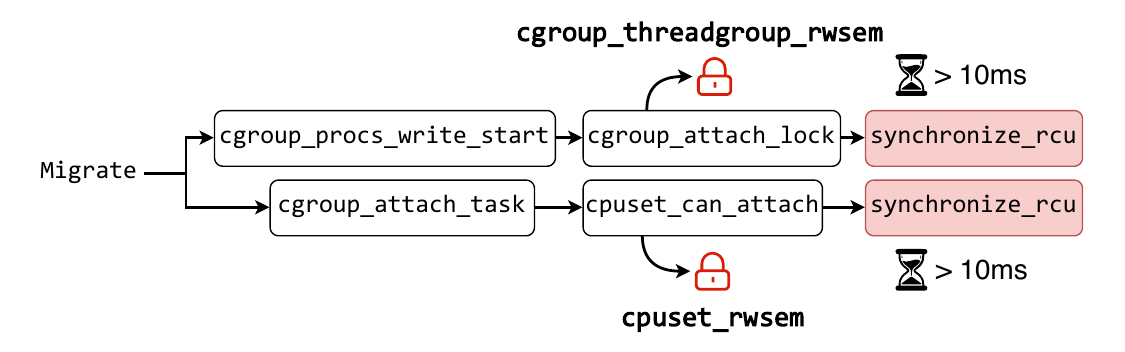}
    \caption{\newc{Call graph of the cgroup migration.}}
    \label{fig:cgroup-move-bottleneck}
\end{figure}

As discussed in \S\ref{sec:isolation-challenge}, cgroup migration suffers from significant latency.
To understand the root cause of this latency, we employed ftrace~\cite{bird2009measuring} to analyze the call graph in Linux during cgroup migration.
\newc{Our analysis reveals that the involvement of two global read-write semaphores in the critical path,
as shown in Figure \ref{fig:cgroup-move-bottleneck}, is the main cause.}
The underlying RCU (Read-Copy Update) synchronization, integral to semaphore implementation, is inherently time-consuming as it necessitates waiting for a grace period to ensure successful write locking.

\systemname employs the \texttt{CLONE\_INTO\_CGROUP} feature, a recent Linux enhancement for task creation.
It significantly speeds up the procedure by assigning a specific cgroup to a process at the time of spawning, rather than post-creation.
Now the workflow in \systemname is:
(1) create the cgroup and (2) assign the cgroup directly while spawning container processes.
Since the process is still invisible to other OS components while spawning, the kernel can bypass the costly synchronization in cgroup migration.
Despite its efficiency, this feature has yet to be adopted in mainstream container runtimes~(e.g., runc). 
In our evaluation, it typically takes only 100 to 300 \textmu s.

%% file: content/agent.tex
\newc{
\section{Extending the Design to VM-Based Agent Execution}\label{sec:extend-to-vm}
\update{
Building on the techniques introduced in the previous section, we further extend \systemname to support VM-based serverless platforms, the dominant model in public clouds.
To evaluate the extended system, we take LLM agents as a representative example of emerging serverless workloads.

Our implementation targets KVM-based microVMs, where each hypervisor runs as a host process isolated by Linux namespaces and cgroups.
As with containers, \systemname still maintains a pool of repurposable sandboxes that can be recycled across hypervisors, minimizing the corresponding creation overheads.
Nevertheless, the intricate execution patterns and substantial memory demands of LLM agents introduce challenges that extend beyond the scope of container-oriented optimizations, motivating the system enhancements detailed in the following subsections.
}

\subsection{Challenge of Agents on VM-based Platforms}\label{sec:agent-vm-challenge}

\begin{figure}[ht]
    \includegraphics[width=.6\linewidth]{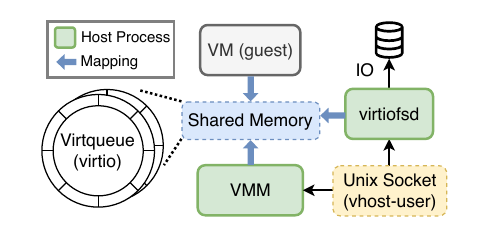}
    \caption{The architecture of virtiofs. VMM means virtual machine monitor.}
    \label{fig:virtiofs-arch}
\end{figure}

As discussed in \S\ref{sec:serverless-agent-limitation}, two primary obstacles hinder cost-efficient serverless execution for agents: low CPU utilization and redundant page cache across guest and host kernels. A common strategy to address low CPU utilization is CPU overcommitment~\cite{ali-overcommit}, which consolidates multiple VMs onto the same set of physical cores.
\update{However, overcommitment introduces two key challenges.}

First, it increases the end-to-end execution latency of agents, particularly for complex agents that rely on browsers. Although the average CPU utilization is low (e.g., only 7\% for ``Game design''), the CPU demand can exhibit significant short-term spikes (e.g., exceeding 40\% for ``Game design'').
Overcommitment exacerbates resource contention among co-located instances. For example, when 200 instances of the ``Game design'' agent are scheduled on 20 physical cores, the average execution latency increases by 25\% (from 107 seconds to 134 seconds).

Second, overcommitment amplifies the impact of redundant page caches. As the number of concurrently running VMs increases, so does the degree of memory redundancy.
Recent hypervisors~\cite{zijun_rund_2022} employ virtiofs~\cite{virtiofs2025}, a shared virtualized filesystem, to mitigate this issue.
\update{
However, we find virtiofs challenging to integrate with our mm-template.
As illustrated in Figure \ref{fig:virtiofs-arch}, the virtiofs delegates guest file operations to a separate userspace daemon called virtiofsd. This architectural decision is explicitly made to enhance security, preventing a compromised guest from exploiting shared file system interfaces to attack the host.
About its data plane, the virtiofsd must access the descriptors in the virtqueue and buffers in the guest memory to serve I/O requests. To ensure performance, the VMM shares the guest memory with virtiofsd rather than copying data between each other.
Unfortunately, this memory sharing model is fundamentally incompatible with our design, which relies on private memory mappings to enable safe copy-on-write semantics across instances.
}

\subsection{Browser Sharing}\label{sec:browser-share}
To mitigate the execution time overhead introduced by overcommitment, \systemname amortizes resource usage across multiple agents.
Many complex agents rely on web browsers to render content or interact with websites. \update{However, browsers are inherently memory- and CPU-intensive, often spawning multiple processes and some agents even needs graphic display (i.e., headed browsers).}
Following the same sharing principle underlying \systemname's design, we extend resource sharing to include browser instances.
\systemname allows multiple agents (e.g., ten) to concurrently share a single browser instance, with each agent operating within its own set of tabs.

\update{
Technically, most agents interact with browsers via automation frameworks such as Playwright~\cite{playwright}, which expose browser operations through language-specific APIs (e.g., Python or JavaScript).
Taking Chromium as an example, these tools communicate with the browser through the Chrome DevTools Protocol (CDP) over WebSocket, sending commands such as opening a new webpage. When CDP is enabled, the browser exposes a WebSocket endpoint that supports multiple concurrent clients.
\systemname launches several browser instances in separate sandboxes and enables multiple agents to connect to the same instance through a shared CDP endpoint.
Because CDP is widely supported, this integration is straightforward. For instance, the Browser-Use framework employed by our ``Shop Assistant'' agent~(Table \ref{tab:agent-chara}) already allows users to specify a CDP endpoint.
Since many browser components, such as network and storage subprocesses, can be multiplexed internally, this design reduces redundant resource consumption and alleviates CPU contention among co-located microVMs.
In the current implementation, we focus on scenarios where agents do not process sensitive user data. Therefore, we do not impose additional security isolation between agent connections sharing a browser instance. Security implications are discussed further in \S\ref{sec:security-limit}.
}

\subsection{Page Cache Mitigation}\label{sec:page-cache-mitigate}
To address the issue of page cache duplication, we reexamine the storage architecture. As discussed in \S\ref{sec:agent-vm-challenge}, the virtiofs-based design in RunD is incompatible with copy-on-write memory sharing on CXL.
\systemname employs the same block device model as Firecracker, allowing storage emulation to be integrated within the hypervisor itself and removing the need for shared memory.
To bypass the guest page cache, we use the virtio-pmem device, a virtualized, byte-addressable persistent memory.
However, simply assigning a dedicated virtio-pmem device to each VM introduces two drawbacks. First, each device occupies substantial host storage (typically tens of gigabytes), as it must store the full VM filesystem. Second, since Linux maintains the page cache at the file level, identical files across different devices result in \emph{duplicated host page caches}.

\begin{figure}[ht]
    \centering
    \includegraphics[width=0.6\linewidth]{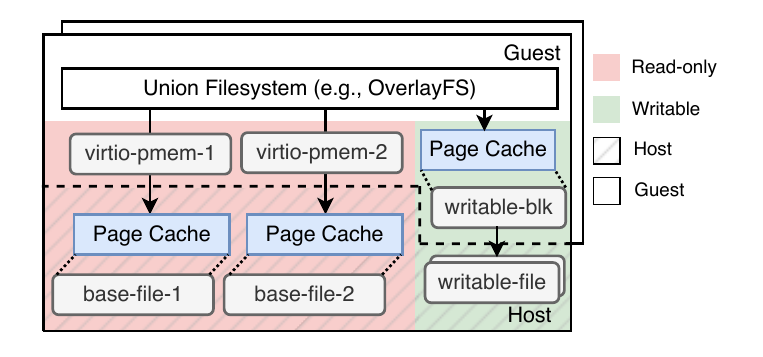}
    \caption{The method to mitigate duplicated page cache.}
    \label{fig:vm-fs}
\end{figure}

To overcome this, \systemname employs a union filesystem strategy, illustrated in Figure \ref{fig:vm-fs}. Each VM is provisioned with two types of storage devices: a read-only base device containing shared files (e.g., libc, browsers), and a writable device for recording per-VM file modifications. The base device is shared among all VMs and mapped as a read-only virtio-pmem device, ensuring a single copy is cached in the host and bypassing the guest page cache.
The writable device is exclusive to each VM and opened with the \texttt{O\_DIRECT} flag in the hypervisor, thereby avoiding host page cache duplication. These two types of devices are combined using a union filesystem (e.g., overlayfs) within the guest, enabling efficient storage isolation and compatibility with the copy-on-write.

}

%% file: content/implementation.tex
\newc{
\section{Implementation}
We have implemented the mm-template mechanism in \systemname based on Linux v6.1, with support for both CXL and RDMA memory pools. The functionality is exposed to users via a pseudo-device under \texttt{/dev}, using a set of \texttt{ioctl} interfaces as illustrated in Figure \ref{fig:mm-template-api}.
Each mm-template is assigned a unique identifier upon creation, which must be provided for subsequent operations. Internally, all templates are managed using an XArray, indexed by their identifiers. The implementation consists of approximately 3,500 lines of code (LoC) modifications to the kernel and 700 lines for the RDMA memory server.

To optimize existing checkpoint/restore (C/R) workflows, \systemname introduces both mm-template and repurposable sandbox mechanisms into CRIU. Rather than reconstructing the mount namespace and cgroup from scratch, \systemname enables joining the existing environment within its sandbox pool. The memory remapping and data copy typically involved in CRIU restore operations are replaced by the efficient \texttt{mmt\_attach} call. Furthermore, we extend CRIU to support the transformation of traditional CRIU checkpoints into mm-templates. This integration involves approximately 2,900 LoC modifications to CRIU.


\systemname integrates CRIU with faasd~\cite{faasd}, a widely-used serverless computing platform, to support container-based function execution. The management of sandbox pools, function-specific overlayfs, and cgroups is delegated to the serverless platform, which possesses sufficient application-level context. For function scheduling, \systemname adopts a simple yet effective policy: if available sandboxes exist in the pool, new function instances are launched via repurposing; otherwise, \systemname falls back to cold start or vanilla CRIU-based startup.

We also extend \systemname to support a VM-based agent execution platform, enabling both lightweight agents for rapid startup and complex agents for high-density workloads. Our survey revealed a lack of open-source microVMs meeting all of \systemname's requirements. Specifically, efficient C/R support is often missing, e.g., Kata Containers lacks native C/R support for microVMs, and gVisor requires costly memory copying.
Moreover, support for DAX (Direct Access) to bypass the guest page cache is essential, which is not supported in Firecracker.
Although DAX feature is standardized in the virtiofs specification, its official implementation \cite{virtiofsd2025} has removed support due to instability in the memory mapping command within the vhost-user protocol.
To address these limitations, we enhance Cloud Hypervisor (CH) to satisfy all requirements. While CH supports virtio-pmem, its default restore mechanism relies on full memory copies. Inspired by the mm-template design, we extend CH to restore memory state via \texttt{mmap} from either a modified DAX device or a regular memory image file.
The VM-based platform implementation consists of 6,000 LoC, with CH enhancements requiring an additional 100 lines.


}

%% file: content/discussion.tex
\section{Discussion}

In this section, we further discuss the security and deployment costs of \systemname.
\begin{enumerate}[label={(\arabic*)}]
    \item Security: \systemname involves reusing or repurposing certain container components, as outlined in Table \ref{tab:container-components}. It is essential to assess the extent to which this reuse may introduce potential security vulnerabilities.
    \item Deployment Costs: \systemname optimizes memory restoration and utilization by leveraging emerging hardware technologies (e.g., CXL and RDMA). This approach presents a tradeoff between the benefits gained and the additional deployment costs incurred.
\end{enumerate}

\subsection{Security}\label{sec:discuss-security}
The mm-template API is implemented through a set of \texttt{ioctl} calls on a pseudo-device driver. To ensure its usage is controlled, only users with root privileges can access that device.
For repurposable sandboxes, the design aims to provide a level of security and isolation that is equivalent to or exceeds that of existing container-based systems.
\update{
In the following, we examine the security implications associated with the reused kernel objects (i.e., container components).
We then analyze other security limitations inherent in the current implementation of \systemname.
}

\subsubsection{Reused Kernel Objects}\label{sec:reused-kernel-objects}
\hfill\\
\textbf{Network namespace}~(netns).
Lightweight containers~(e.g., used in Mitosis~\cite{wei_no_2022}) do not employ netns, leading all instances to share the same network environment.
In contrast, \systemname assigns a separate netns to each instance. During repurposing, the previously opened network connections will be forcibly terminated to prevent data leakage.
However, certain states, including the network configurations~(e.g., firewall rules and routing tables) and statistics~(e.g., received bytes of the veth interfaces), are preserved.
Many serverless functions, including all evaluated in our study, do not modify network configurations. 
Therefore, these residual states do not expose data generated during function execution.
\update{
For other functions that customize the network or ask for extreme security, we need fallback to create absolutely clean network environment from scratch as usual.
}



\noindent \textbf{Rootfs}.
Each rootfs contains a mount namespace and a union filesystem. Its security implications has been discussed in \S\ref{sec:rootfs-reconfig}. During repurposing, \systemname first kills the processes, and then purges the file modifications of the previous instance. Thus, \systemname does not leak any data, including memory or files, produced by the processing of the last function.

\noindent \textbf{Cgroup}.
The reuse of cgroup is confirmed and adopted by production systems like Rund~\cite{zijun_rund_2022}. Still, \systemname provides a cgroup per instance, as standard Docker containers.

A ``share'' strategy has been proposed by Pagurus~\cite{li_help_nodate}. A pool of zygote containers is maintained to mitigate cold start overhead.
Nevertheless, all child containers forked from the same zygote container share the isolated components~(e.g., rootfs and netns), resulting in poorer isolation compared with \systemname.

\subsubsection{Security Limitations}\label{sec:security-limit}
Although \systemname prevents data leakage and enforces the proper level of isolation during repurposing, there are other potential security limitations.
\begin{enumerate}[label={(\arabic*)}]
    \item ASLR (Address Space Layout Randomization). 
    In \systemname, all restored container instances share the same memory layout (e.g., the virtual address of the stack) as the mm-template they are attached during restoring, preventing ASLR from introducing randomness.
    This issue is prevalent in all Checkpoint/Restore and fork-based schemes, where restored or forked containers inherit the virtual address layout of the snapshot or parent process.
    \item Side-channel attacks related to memory deduplication across functions~\cite{yarom_flushreload_nodate}. A possible solution is to only enable it for functions from the same users.
    \item \update{
    Unintentional data sharing (e.g., cookies) among browser-sharing agents. A possible solution is integrating advanced isolation mechanisms of modern browsers, like Firefox Containers~\cite{firefox_container}, which provide privacy separation between tabs within a single browser instance.%
    }%
    \item Data protection during transfer. For CXL, the 2.0 specification propose security features, such as IDE~(Integrity and Data Encryption). For RDMA, it is possible to encrypt the memory images during transfers.
\end{enumerate}

\subsection{Deployment Cost}\label{sec:deployment-cost}
For RDMA-based deployment, there is no additional cost since cloud providers already offer RDMA devices~(e.g., eRD-\allowbreak MA~\cite{cao2024efficient}, EFA~\cite{ziegler_efa_2022}) in production.
For CXL-based deployment, a quantitative cost analysis is currently not feasible because CXL 2.0 switches are still in the demo phase. However, we are confirmed from manufacturers that memory expanders’ and switches’ price will be comparable to DDR5 DIMMs and IB switches, respectively.
Previous works\allowbreak~\cite{berger_design_2023} showed that the cost of switches and multi-headed memory controllers are within 5\% of the original servers, at the scale of 16 sockets.
Thus, we anticipate rack-level deployment of CXL memory pools will be available at an acceptable cost in the near future, motivating our work.
A rack-level mini-cluster with around 10 machines and 20TB memory would be sufficient to capitalize on the benefits of \systemname. 

Regarding the energy cost for occupying DRAM and the comparison with pre-warm/keep-alive approaches, \systemname reduces the overall memory footprint by enabling cross-machine-intra-rack deduplication. Only one copy is needed per rack if it is read-only, reducing the cost by a factor of the number of machines (\textasciitilde 10). In contrast, each kept-warm container requires an independent copy, leading to excessive duplication and memory costs. 
For larger clusters, it is possible to blend CXL~(intra-rack) and RDMA~(inter-rack), by adjusting the mm-template, to further enhance the scalability.

%% file: content/eval.tex
\section{Evaluations}
\newc{
In this section, we present the experimental evaluation of \systemname. \S\ref{sec:eval-methodology} to \S\ref{sec:cxl-vs-rdma} focus on container-based platforms, assessing the effectiveness of repurposable sandboxes and mm-template.
\S\ref{sec:agents-eval} extends the evaluation to VM-based agent platforms, highlighting \systemname's enhancements for browser sharing and mitigation of duplicated page cache.
}

\subsection{Methodology}\label{sec:eval-methodology}
\newcommand{\ecxl}{T-CXL\xspace}
\newcommand{\erdma}{T-RDMA\xspace}


\noindent \textbf{Testbed}.
We conducted evaluations of \systemname using memory pools based on both CXL and RDMA technologies,
termed \ecxl and \erdma in the following sections, respectively.
The test platform was equipped with dual 32-core Intel Xeon Gold 6454S CPUs, 256 GB of RAM, and a 7 TB Samsung PM9A3 SSD. Additionally, the platform was connected to a 128 GB experimental Samsung CXL memory device and a Soft-RoCE RDMA device. 
In our tests, the latency for accessing remote memory was 641.1 ms for CXL and 6 \textmu s for RDMA.


\begin{table}[b]
\caption{The evaluated functions. The last column shows the number of threads that need to be restored.}
\label{tab:evaluated-funcs}
\begin{tabularx}{\linewidth}{p{.06\linewidth}|p{.09\linewidth}p{.42\linewidth}p{.14\linewidth}p{.13\linewidth}}
\toprule
Func & Lang & Description                     & Mem Size & \# Thread \\ \midrule
DH   & Python   & Dynamic web pages generating.   & 50.4 M    & 14        \\
JS   & Python   & Deserialize and serialize json. & 94.9 M    & 14        \\
PR   & Python   & Pagrank algorithm.              & 116 M     & 395       \\
IR   & Python   & Deep learning (ResNet).         & 855 M     & 141       \\
IP   & Python   & Image rotating and flipping.    & 67.1 M    & 15        \\
VP   & Python   & Gray-scale effect on video.     & 324 M     & 204       \\
CH   & Python   & HTML tables rendering.          & 94.9 M    & 38        \\
CR   & Nodejs   & AES encryption algorithm.     & 124 M     & 16 \\  
JJS  & Nodejs   & Similar to JS.                & 111 M     & 21        \\ 
IFR  & Nodejs   & Similar to IP.                & 253 M     & 21        \\
\bottomrule
\end{tabularx}
\end{table}

\noindent \textbf{Evaluated Functions}.
Our method is universally applicable across diverse language runtimes and is able to handle multi-threaded and multi-process scenarios.
Thus, for our evaluation, we selected a wide range of applications from SeBS \cite{copik_sebs_2021} and Function-Bench \cite{kim_functionbench_2019} (Table \ref{tab:evaluated-funcs}).
Given the prevalent use of Node.js and Python by AWS Lambda developers\cite{state-of-serverless}, 
we also ported several Python (Py) functions to Node.js (Node).
This further highlights \systemname's ability to repurpose between heterogeneous languages.

\noindent \textbf{Baselines}.
\systemname is based on \underline{faasd}, a widely used serverless platform that not only provides a baseline for our comparative analysis but also supports \underline{CRIU} natively.

Additionally, the current predominant approach to mitigating cold start issues in serverless computing is ``lazy restoration'', which prioritizes the recovery of essential states and delays other tasks (e.g., memory content restoration) until necessary.
Thus, we also compare \systemname with  \underline{REAP} and \underline{FaaSnap}, two state-of-the-art lazy restoration methods that utilize Firecracker. 
Notably, FaaSnap builds on top of REAP by introducing an asynchronous prefetch policy.
Each guest for FaaSnap and REAP has 2 vCPUs and 2 GB memory.

During our evaluation, we noted that the network configuration phase during Firecracker initialization introduces significant overhead, which can reach up to 600 ms under scenarios of high load and concurrency. 
This overhead allows \systemname to outperform them substantially.
To facilitate a more meaningful comparison, we developed a network namespace pool for both REAP and FaaSnap. 
After a VM is terminated, its network environment is recycled into this pool for later reuse, analogous to our repurposable sandbox pool. 
We refer to these enhanced implementations as \underline{REAP+} and \underline{FaaSnap+}, respectively.

\noindent \textbf{Schedule Policy}.
We implement a widely used scheduling policy across all evaluated methods. After the invocation, instances are retained in a container pool for a fixed duration (e.g., 10 minutes) akin to Openwhisk \cite{openwhisk-warmed-container}, commonly referred to as keep-alive.
This approach allows for the immediate reutilization of cached instances for new invocations of the \emph{same function}. 
The container pool works as a Least Recently Used (LRU) list, organized by the most recent activity.


\noindent \textbf{Workloads}.
We assessed the performance of \systemname using both synthetic and real-world workload traces. 
Specifically, bursty loads and diurnal patterns are the two most commonly observed patterns in real-world serverless platforms 
that lead to load instability and thus diminish the effectiveness of the keep-alive strategy~\cite{shahrad_serverless_nodate,fuerst_faascache_2021,roy_icebreaker_2022,li_help_nodate,wang_faasnet_nodate,singhvi_atoll_2021}. 
To accurately simulate these conditions, we designed two specific workloads: W1 and W2. 
W1 replicates bursty traffic patterns, with intervals between consecutive bursts \emph{longer than the keep-alive threshold}. 
In contrast, W2 emulates diurnal traffic fluctuations, cycling through various functions under \emph{tight memory limits} (a soft memory cap of 32GB is applied in W2, compared to the 64GB used in other tests). 
To further substantiate \systemname's utility in practical scenarios, we also incorporated industry traces from Azure~\cite{shahrad_serverless_nodate} and Huawei~\cite{joosen_how_2023} into our evaluations.

To ensure the effectiveness of traditional caching mechanisms, all tests include a warm-up phase of about 5 minutes. 
Additionally, snapshot images of CRIU, REAP, and FaaSnap are stored on a CXL-memory-backed tmpfs to eliminate disk overhead, ensuring a fair comparison with \ecxl.

In the following evaluation, we focus on the following aspects.
\begin{enumerate}[label={(\arabic*)}]
    \item \textbf{Performance}~(\S\ref{sec:representative-workload}, \S\ref{sec:real-world-workload}, \S\ref{sec:eval-startup-agents}, \S\ref{sec:eval-browser-share-agents}): What is the benefit of \systemname on \emph{end-to-end} latency, especially under concurrency and the P99 latency?
    \item \textbf{Memory utilization}~(\S\ref{sec:representative-workload}, \S\ref{sec:real-world-workload}, \S\ref{sec:eval-mitigate-page-cache-agents}): What is the benefit of \systemname on the reduction of \emph{memory usage}?
    \item \textbf{Optimization Breakdown}~(\S\ref{sec:opt-breakdown}): What is the contribution of each optimization proposed by \systemname?
    \item \textbf{Memory Pool Comparison}~(\S\ref{sec:cxl-vs-rdma}): What is the difference between CXL and RDMA memory pool?
\end{enumerate}

\subsection{Representative Workload}\label{sec:representative-workload}

\begin{figure}[tb]
    \centering
    \begin{subfigure}{.98\linewidth}
        \includegraphics[width=\linewidth]{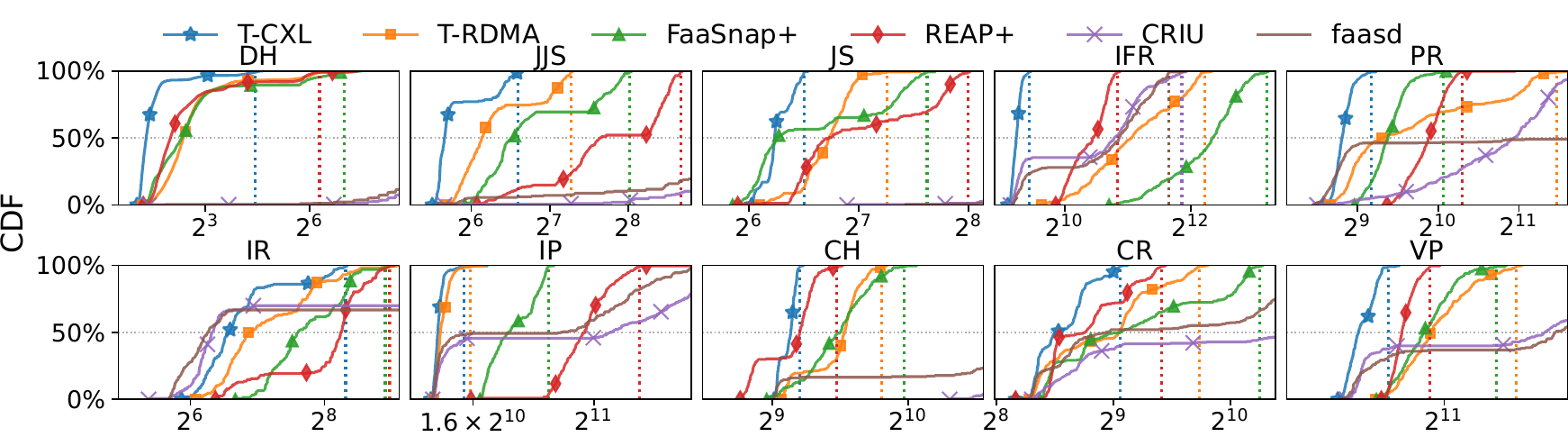}
        \caption{W1: bursty loads with intervals longer than the keep-alive threshold.}
        \label{fig:cdf-w1}
    \end{subfigure}
    \begin{subfigure}{\linewidth}
        \includegraphics[width=\linewidth]{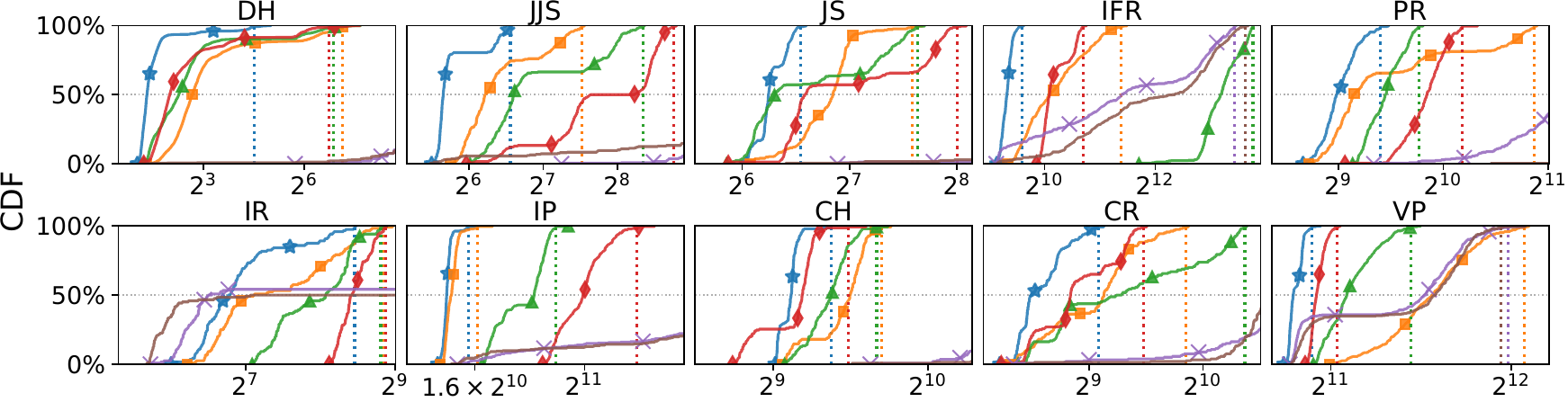}
        \caption{W2: diurnal and heavy traffic under tight memory limits.}
        \label{fig:cdf-w2}
    \end{subfigure}
    \caption{CDF of different functions' E2E latency under two representative workloads, W1 and W2. The vertical dotted line indicates the P99 E2E latency. Note that x-axis denotes the E2E latency in milliseconds on a logarithmic scale with a base of 2.}
    \label{fig:representative-workload-cdf}
\end{figure}

Figure \ref{fig:representative-workload-cdf} displays the cumulative distribution function (CDF) of end-to-end (E2E) latency for functions within the W1 and W2 workloads, capturing over 4k invocations in 30 minutes. 
For clarity,  the x-axis shows the logarithmic scale (base 2) latency, and the results for CRIU and faasd are partially truncated due to their relatively longer latencies.

As a summary, in terms of tail latency, \ecxl consistently outperforms all other solutions, achieving a speedup ranging from 1.11$\times$-5.69$\times$~(1.17$\times$-18$\times$) for P99 latency and up to 5.9$\times$~(8.6$\times$) for median latency, compared with REAP+~(FaaSnap+).
The superior performance primarily stems from  \ecxl's reduced execution time, as FaaSnap+ and REAP+ incur more context switches and on-demand restoration overheads, particularly under conditions of high concurrency. 
\erdma also shows promising performance; however, it does not perform as well in certain scenarios due to the higher latency associated with RDMA access compared to the CXL-backed tmpfs used in REAP+ and FaaSnap+.

\begin{figure}[tb]
    \centering
    \includegraphics[width=.6\linewidth]{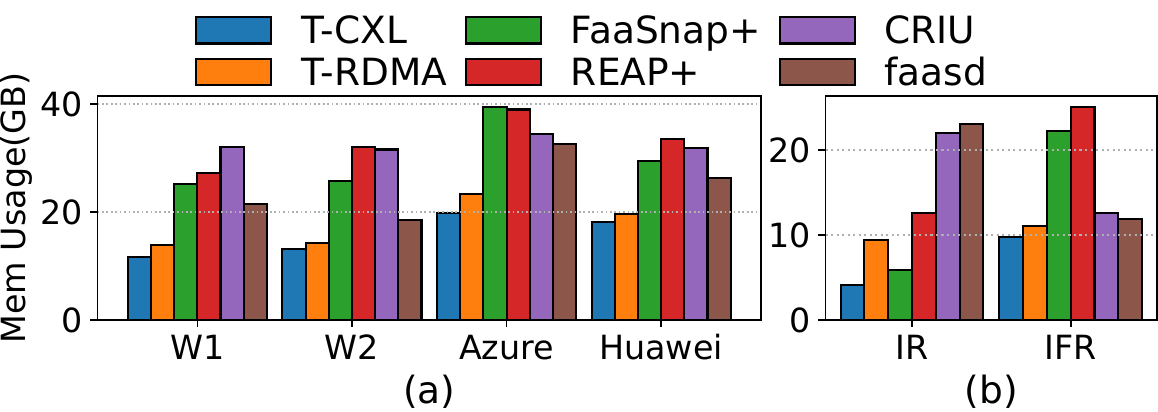}
    \caption{\update{(a) Peak memory usage during four workload tests. (b) The memory usage when starting 50 instances of IR and IFR, respectively.}}
    \label{fig:trace-mem-usgae}
\end{figure}

\begin{figure}[tb]
    \centering
    \includegraphics[width=.6\linewidth]{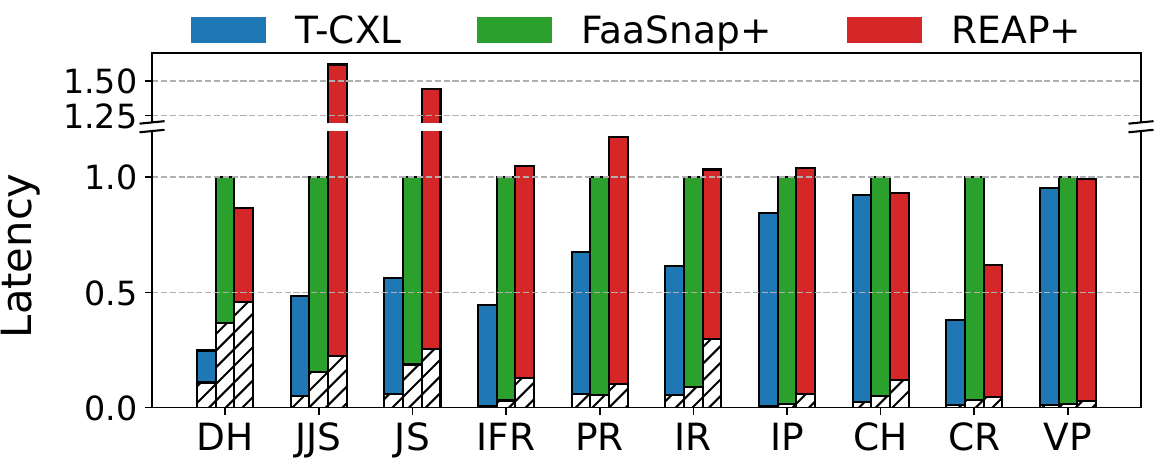}%
    \caption{\update{Normalized E2E latency without concurrency, while the hatched region denotes the startup time.}}
\label{fig:cold-start-vs-reap-faasnap}
\end{figure}

Regarding memory utilization, as detailed in Figure \ref{fig:trace-mem-usgae}a,
\ecxl achieves an average reduction in memory usage relative to faasd, CRIU, REAP+, and FaaSnap+ in W1 and W2 by 37.4\%, 61.2\%, 58.2\%, and 51.5\%, respectively.
\erdma exhibits a comparable level of memory savings.
These improvements are attributed to memory sharing, deduplication, and reduced number of instances to keep in memory, thanks to our repurposing technique that allows for the effective share of execution environments across various functions.

\subsubsection{Comparing with CRIU and faasd}\label{sec:eval-vs-CRIU-faasd}
The suboptimal performance of CRIU and faasd stems from their high startup latency. 
As given in Table \ref{tab:container-components},
creating isolation environments can take over one second
under conditions of heavy loads and high concurrency. 
Additionally, faasd encounters delays due to application initialization during cold starts, whereas CRIU's restoration process is hampered by the expensive copy-based memory restoration. 
For instance, launching a CR instance takes 1.7 s at P99, whereas its execution time is only about 500 ms.
Thanks to our repurposable sandboxes and mm-templates, \systemname significantly reduces the costs associated with isolation environment creation and memory restoration. 
It takes merely 15 milliseconds for \ecxl to start the same CR instance at P99, representing a reduction of more than 100$\times$. 
In \S\ref{sec:opt-breakdown}, we will explore the individual contributions of the repurposing and mm-template techniques.


\ecxl relies on CXL, which has higher latency than local DRAM, resulting in degraded execution performance.
This explains CRIU's better P50 IR performance in Fig \ref{fig:representative-workload-cdf}. 
For instance, \ecxl nearly doubles the execution time of DH and IR due to their short runtimes (<100 ms), while other functions see about a 10\% increase on average. 
However, during cold starts, \ecxl's E2E latency is significantly lower than CRIU's, thanks to its efficient startup process.
Additionally, performance can be improved by configuring mm-templates to store hot regions of memory image in local DRAM.

\subsubsection{Comparing with REAP+ and FaaSnap+}\label{sec:eval-vs-reap-faasnap}
Both REAP+ and FaaSnap+ are based on Firecracker. While the primary focus is on enhancing security, the additional layer of the hypervisor facilitates efficient state restoration.
Despite the reduced gap, \systemname, which is container-based, still achieves better start latency. 
For example, it takes only 13 ms to start an instance of CH at P99, compared to 49 (90) ms for FaaSnap+ (REAP+). 
Moreover, Firecracker leads to significantly higher memory consumption (as shown in Figure \ref{fig:trace-mem-usgae}a) than containers, due to each guest OS maintaining exclusive resources, such as page caches.
A microbenchmark,  illustrated in Figure \ref{fig:trace-mem-usgae}b, further highlights this issue.
When starting 50 function instances, we observed that FaaSnap and REAP even doubled the memory usage compared to \ecxl.


In addition to startup latency, \systemname achieves lower tail latency than both REAP+ and FaaSnap+ due to its much more stable and {\bf shorter execution time}.
REAP and FaaSnap use a lazy restore approach that only delays, rather than eliminates, restoration overheads during execution, especially for memory.
For instance, handling each page fault still requires several microseconds by the OS, even when their snapshots are stored on a CXL-based tmpfs. 
In contrast, \systemname eagerly restores most states instead of on-demand, such as opened files.
Further, by leveraging CXL's byte addressability, mm-template avoids page faults for read-only pages.
According to our investigations, about 24\% to 90\% of the pages used during execution are read-only. 
Thus, as illustrated in Figure \ref{fig:cold-start-vs-reap-faasnap}, the execution time for \systemname is considerably shorter than that for REAP+ and FaaSnap+.

\subsubsection{The effect of function characteristics}
Different applications exhibit distinct characteristics that influence their performance. The functions we evaluated can be categorized into three groups.
\begin{enumerate}[label={(\arabic*)}]
    \item Memory-insensitive applications, such as compute-intensive (like VP and IP) and I/O-intensive (like CH) applications. The latency for these applications does not show a significant disparity among different methods, as they are primarily CPU-bound or I/O-bound rather than memory-bound. 
    As a result, the differences in E2E latency between CXL, RDMA, and even userfaultfd-based page faults (used in REAP) are not evident.
    \item Applications with a large memory footprint and complex memory access patterns, including IR, PR, and IFR. They lead to more minor page faults in REAP+ and FaaSnap+, resulting in longer execution time than \ecxl. \erdma can also experiences longer tail latency, especially during burst, due to unstable P99 latency of RDMA under high request rates.
    \item Applications with brief execution time, which is common in serverless scenarios, including DH, JS, CR and JJS. Any extra overhead, such as minor page faults in FaaSnap or REAP, and longer latency for RDMA, significantly impacts their E2E latency. The shorter startup latency and zero cost for read accesses in \ecxl demonstrate more pronounced effects for these functions.
\end{enumerate}

\begin{figure}[tb]
    \centering
    \begin{subfigure}{.65\linewidth}
        \includegraphics[width=\linewidth]{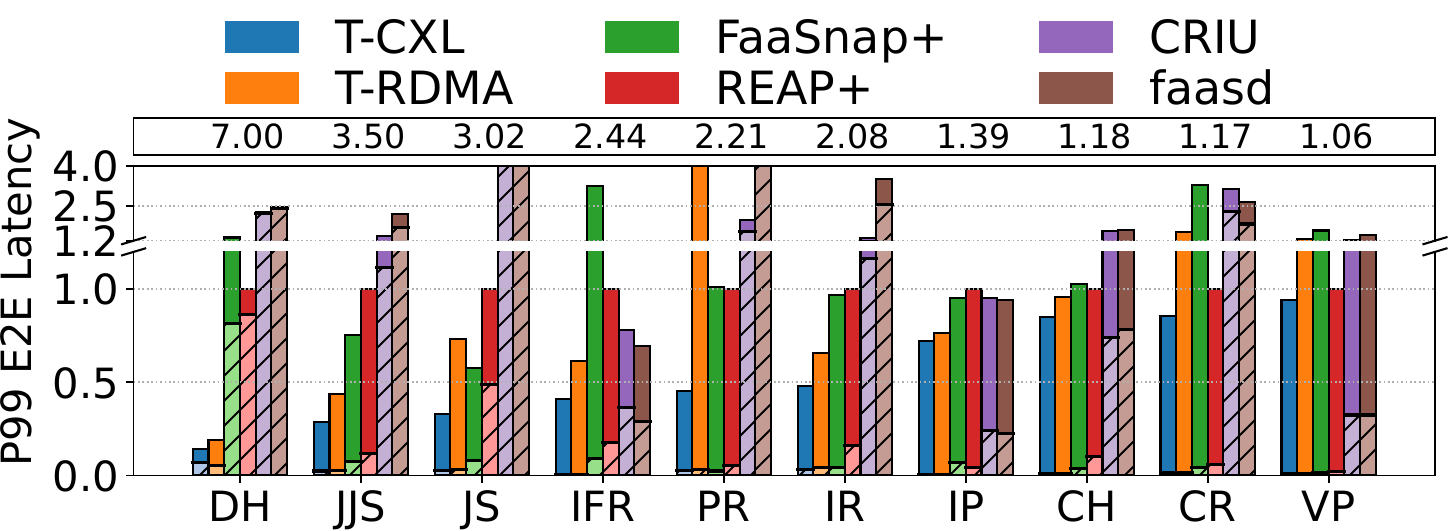}%
        \caption{Azure workload}
    \end{subfigure}
    \begin{subfigure}{.64\linewidth}
        \includegraphics[width=\linewidth]{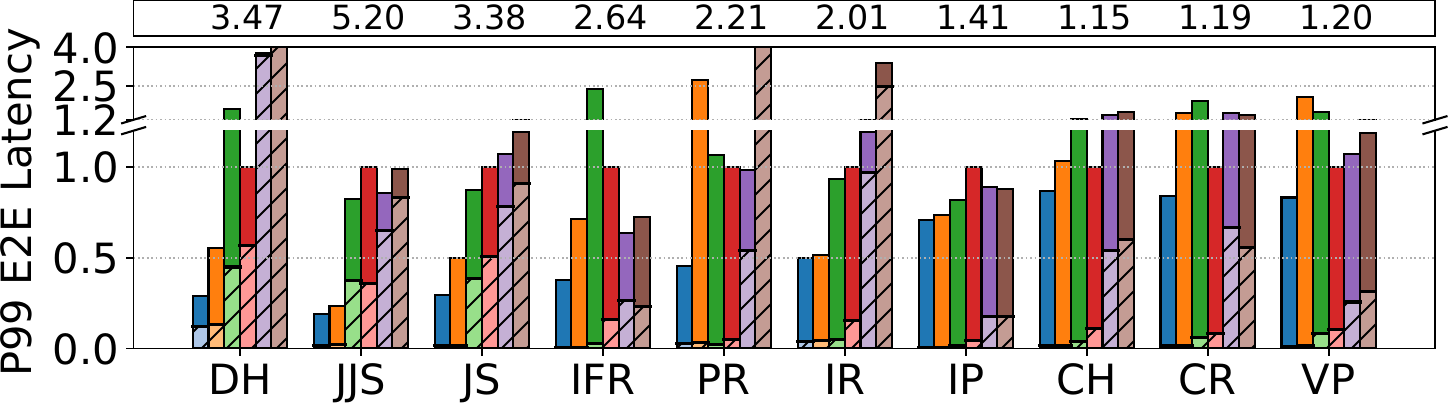}%
        \caption{Huawei workload}
    \end{subfigure}%
    \caption{\update{P99 E2E latency for read-world workloads, normalized against REAP latency. Each bar is divided into two segments: the upper un-hatched represents execution time, and the lower hatched indicates startup time. Top numbers are the speedup of \ecxl compared to REAP+.}}
    \label{fig:real-trace-p99-lat}
\end{figure}

\subsection{\update{Real-world Workload}}\label{sec:real-world-workload}

To further justify the effectiveness of our designs, we tested \systemname under two industrial and complex workloads (Figure \ref{fig:real-trace-p99-lat}).
As both datasets only record the number of invocations per minute,
we randomly distributed those within each minute, with a probability of creating skew or bursty loads to imitate real-world conditions.


In summary, for tail latency, \ecxl achieves speedups ranging from 1.06$\times$-7.00$\times$ and 1.16$\times$-9.25$\times$ compared with REAP+ and FaaSnap+, due to our shorter execution time under heavy loads.
\erdma falls behind REAP+ and FaaSnap+ in JS, VP, CH, CR, and PR. 
The unstable P99 memory access latency of RDMA still incurs a delay in these heavy-load scenarios.
Nevertheless, \erdma still achieves a speedup of 1.29$\times$-4.28$\times$ and 1.11$\times$-4.64$\times$ against REAP+ and FaaSnap+ for other functions.
As shown in Figure \ref{fig:trace-mem-usgae}, \ecxl and \erdma
reduce memory usage by over 25\% compared to all baselines in both workloads.
Specifically, \ecxl reduces memory usage by up to 49\% relative to REAP+ and FaaSnap+.
\erdma consumes, on average, 10\% more memory than \ecxl.

\subsection{Breakdown of Optimization Steps}\label{sec:opt-breakdown}
To delve deeper into the contribution of each optimization, we analyzed the E2E latency by enabling different optimizations in \systemname step by step.


\begin{figure}[tb]
    \centering
    \includegraphics[width=.62\linewidth]{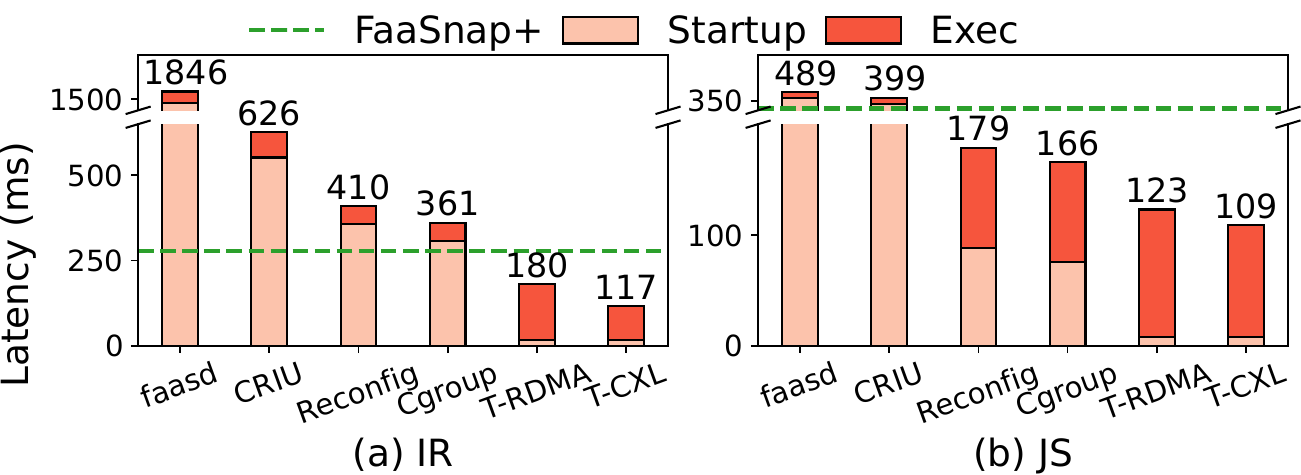}
    \caption{\update{Optimization steps and its effects to E2E latency. The green line indicates the E2E latency of FaaSnap+.}}
    \label{fig:opt-breakdown}
\end{figure}

In Figure \ref{fig:opt-breakdown}, the ``Reconfig'' optimization applies sandbox repurposing without cgroup optimization,
reducing startup latency by approximately 200 ms. 
Regarding rootfs reconfiguration, one subtask of the repurpose,
it takes more than 30 ms to restore the rootfs and mount namespace in CRIU.
In contrast, \systemname's reconfiguration process involves only two system calls and typically completes rootfs preparation in less than 1 ms. 
Additionally, the ``Cgroup'' optimization employs the \texttt{CLONE\_INTO\_CGROUP} to bypass synchronization delays within the kernel, further reducing startup latency by 49 ms for IR and 13 ms for JS.

After that, the function's startup latency is predominantly governed by memory restoration (>85\%). 
\systemname reduces this cost through mm-template.
First, it only needs to copy a small amount of metadata rather than the memory pages. For example, the metadata is less than 400 KB, while the memory contents exceed 70 MB for JS.
Second, it eliminates the need for reconstructing the virtual memory layout since it is already preserved in the mm-template, thus avoiding numerous \texttt{mmap()} system calls in CRIU.
The larger the memory footprint, the greater the reduction in startup latency achieved by mm-template.
The mm-template alone reduces startup latency by 290 ms for IR and 67 ms for JS, allowing \systemname to launch an IR and JS instance in just 18 ms and 8 ms, respectively.

Utilizing the mm-template with a remote memory pool leads to a slight increase in execution time compared to when local DRAM is used, because of the inherently longer access latencies associated with CXL and RDMA memory.
Specifically, an additional 24 ms (88 ms) for IR and 11 ms (25 ms) for JS, are observed for \ecxl~(\erdma), compared to CRIU.
Nonetheless, the overall E2E latency still experiences a substantial reduction.

We also evaluated the ``Cgroup''~(i.e., enable repurposable sandboxes but without mm-template) with industrial workloads: \ecxl still outperforms ``Cgroup'' in all functions, achieving the speedup ranging from 1.04$\times$ to 2.32$\times$.

\begin{figure}[tb]
    \centering
    \includegraphics[width=.58\linewidth]{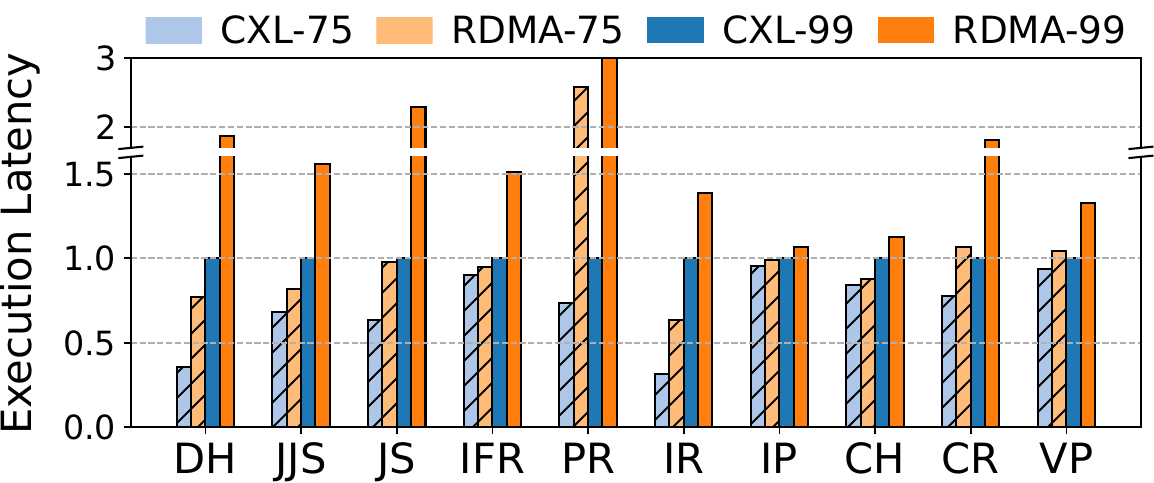}
    \caption{\update{Normalized execution latency of \ecxl and \erdma. The hatched bar indicates P75 latency, while the un-hatched bar indicates P99 latency.}}
    \label{fig:lat-cxl-vs-rdma}
\end{figure}

\subsection{\systemname-CXL vs. \systemname-RDMA}\label{sec:cxl-vs-rdma}
The flexibility of \systemname enables it to leverage various interconnect technologies. 
However, the differing physical characteristics of CXL and RDMA lead to disparities in execution time.
As shown in Figure \ref{fig:lat-cxl-vs-rdma}, \ecxl outperforms \erdma across all functions, with speed increases ranging from 1.04$\times$ to 3.51$\times$ for P75 latency. 
This improvement is attributed to (1) the faster access latency of CXL memory, and (2) the utilization of CXL's byte-addressable feature.
The zero additional software-level cost for read accesses with mm-templates allows \ecxl to eliminate page faults that \erdma encounters for read-only pages. 
Analysis of our evaluated functions reveals that a significant portion of the memory pages are read-only, ranging from 24\% to 90\%. 

Furthermore, the disparity in P99 latency between \ecxl and \erdma is even more pronounced than at P75. 
Under extreme loads at P99, the heavy RDMA traffic exacerbates CPU load and flow interference, and increases contention for network resources such as the switch~\cite{zhu2015congestion} and RDMA NIC~\cite{chen2019scalable}. 
Several works report similar observations, with this performance cliff being nearly fivefold in instances of burst traffic~\cite{kong2023understanding}.
In contrast, CXL memory offers higher IOPS and more stable latency at P99. 
CPU usage monitored during two industrial tests with \texttt{mpstat} shows that \erdma's total CPU usage is 1.24$\times$ and 1.23$\times$ higher than \ecxl for Azure and Huawei traces, respectively.

Figure \ref{fig:trace-mem-usgae}b also demonstrates that for read-heavy functions like IR, \ecxl primarily accesses remote memory directly, avoiding the allocation of local pages. It substantially saves memory (43.5\%) compared to \erdma. 
Conversely, for write-heavy functions like IFR, \ecxl shows lesser memory gains (13\%) as both trigger COW for write access.

Nonetheless, since all states in the memory pool are read-only, a multi-layered architecture that strategically places hot pages in CXL and cold pages in RDMA,
integrates seamlessly with our approach.
The specific cache eviction strategies are orthogonal to our core implementation.

\newc{
\subsection{The Enhancement of \systemname over VM-based Agent Platform}\label{sec:agents-eval}
\newcommand{\agentB}{E2B\xspace}
\newcommand{\agentBp}{E2B+\xspace}


\noindent \textbf{Evaluated Agents.}
We evaluate a set of representative agents listed in Table \ref{tab:agent-chara}, covering both lightweight and complex designs.
Agent execution exhibits substantial non-determinism. First, identical user inputs may yield different execution paths due to the inherent randomness of LLM outputs. Second, runtime variability in the LLM inference backend may impact execution latency.
To ensure fair and reproducible evaluation, we collect execution traces from real LLM runs, including the exact outputs and corresponding response times. During benchmarking, agents interact with a simulated inference server that replays the recorded outputs and enforces the same response latency. This methodology ensures deterministic execution across repeated runs, enabling reliable comparisons.

\noindent \textbf{Configurations.}
The first three agents in Table \ref{tab:agent-chara} (i.e., Blackjack, Bug Fixer, and MapReduce) are deployed in VMs provisioned with 1 vCPU, 2 GB of memory, and 5 GB of storage. The remaining agents, which rely on a browser runtime, are allocated 4 GB of memory, with other configurations remaining the same.
To evaluate \systemname's optimization effectiveness under overcommitment, we concurrently launch 200 agent instances across 20 physical cores in the following experiments.

\noindent \textbf{Baselines.}
To the best of our knowledge, there is no open-source cloud platform specifically designed for serving LLM-based agents. A closely related paradigm is the code interpreter, where code generated by an LLM is executed in an isolated environment, and the results are returned to the agent.
Representative platforms, such as E2B~\cite{e2b}, commonly leverage microVMs to provide strong isolation and predictable performance. We adopt \agentB as our baseline system. Notably, \agentB already incorporates checkpoint/restore (C/R) techniques to reduce startup latency. To improve its efficiency further, we enhance \agentB with a key modification: the adoption of the root filesystem (rootfs) mapping scheme from RunD. We refer to this enhanced version as \agentBp in the remainder of the evaluation.


\subsubsection{\update{The Startup Efficiency for Agents}.}\label{sec:eval-startup-agents}
Lightweight agents, similar to traditional serverless functions, have short execution durations and are therefore highly sensitive to startup latency. 
\update{
Figure \ref{fig:agent-start-lat} shows the startup latency of all agents listed in Table \ref{tab:agent-chara} across different systems. \systemname reduces startup latency by approximately 35\% $\sim$ 49\% compared to \agentB and 36\% $\sim$ 49\% compared to \agentBp, demonstrating its effectiveness in minimizing launch overhead.
This improvement is primarily attributed to \systemname's efficient creation of isolated sandboxes. For instance, \agentB introduces considerable startup overhead, requiring roughly 15 ms for network environment setup and an additional 10 ms for cgroup migration.

To further assess the benefits of the mm-template mechanism in VM-based environments, we also measure the startup latency of the vanilla Cloud Hypervisor (CH), which restores VM memory states via full memory copying.
This results in latencies ranging from 770 ms to 1700 ms. Agents with larger memory footprints, such as “SA”, “BS”, and “GD”, which are equipped with double-sized memory, require more time for CH to start up compared to other agents.
}
In contrast, \systemname employs mm-template to restore memory using a single system call, avoiding full copying of memory. Instead, memory pages are populated lazily at runtime. This design significantly reduces memory restoration overhead and, in turn, startup latency, making \systemname particularly well-suited for lightweight agent workloads.

\begin{figure}[h!]
    \centering
    \includegraphics[width=0.62\linewidth]{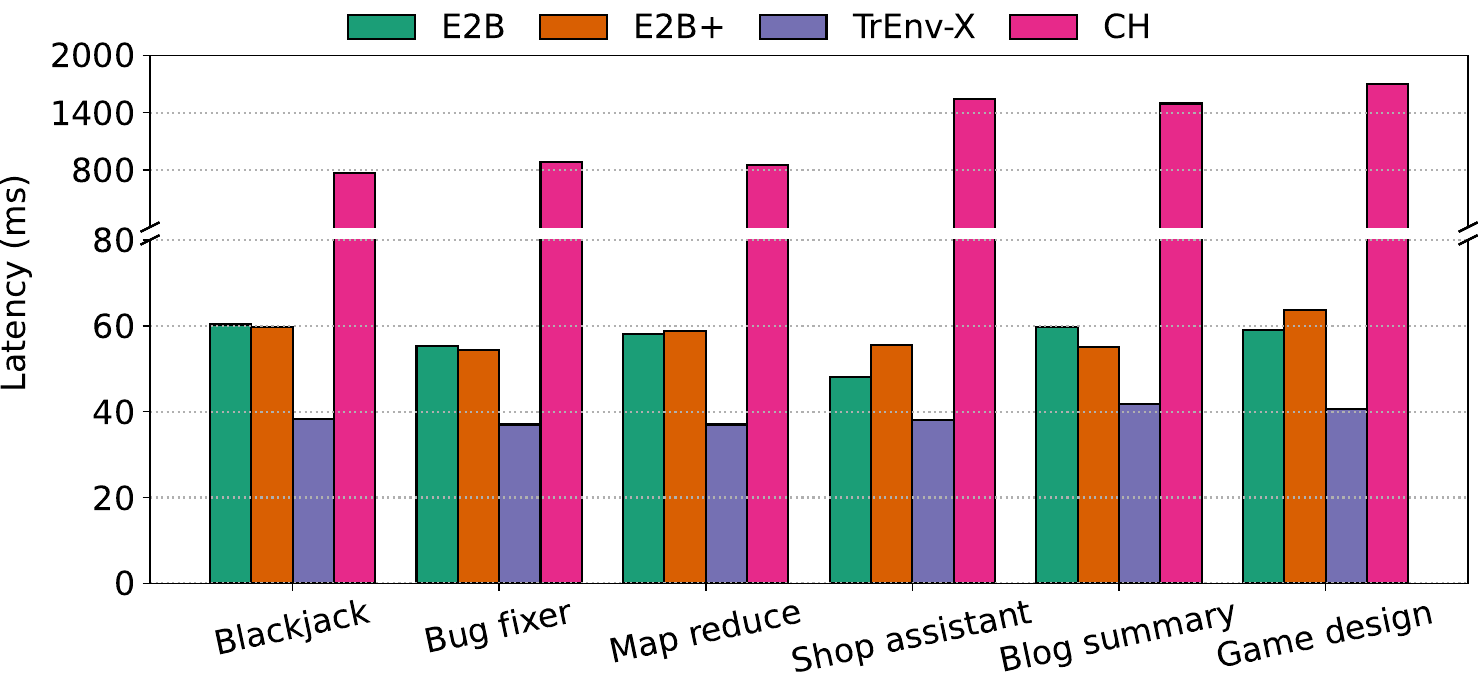}
    \caption{\update{The startup latency of LLM agents in Table \ref{tab:agent-chara}. CH means the vanilla Cloud Hypervisor.}}
    \label{fig:agent-start-lat}
\end{figure}

\subsubsection{Browser sharing.}\label{sec:eval-browser-share-agents}

\begin{figure}[t]
\centering
    \begin{minipage}[t]{0.33\textwidth}
        \centering
        \includegraphics[width=\linewidth]{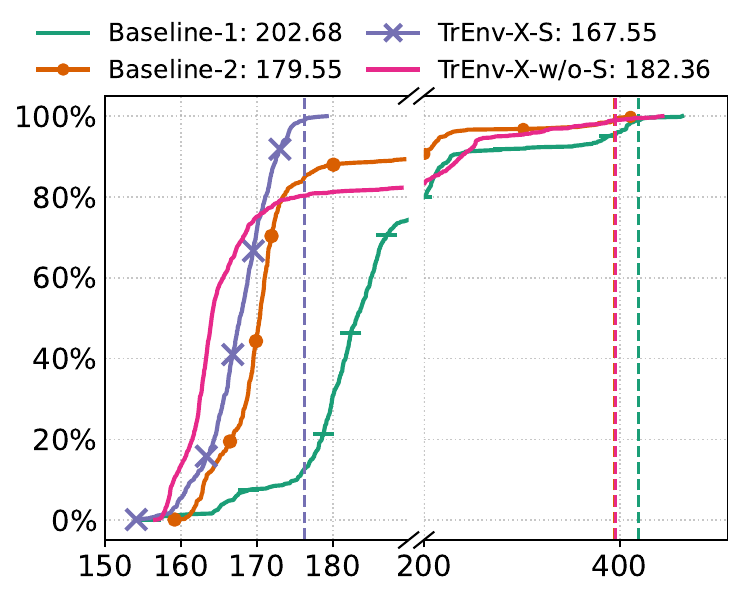}
        \subcaption{Shop assistant.}
    \end{minipage}%
    \hfill
    \begin{minipage}[t]{0.33\textwidth}
        \centering
        \includegraphics[width=\linewidth]{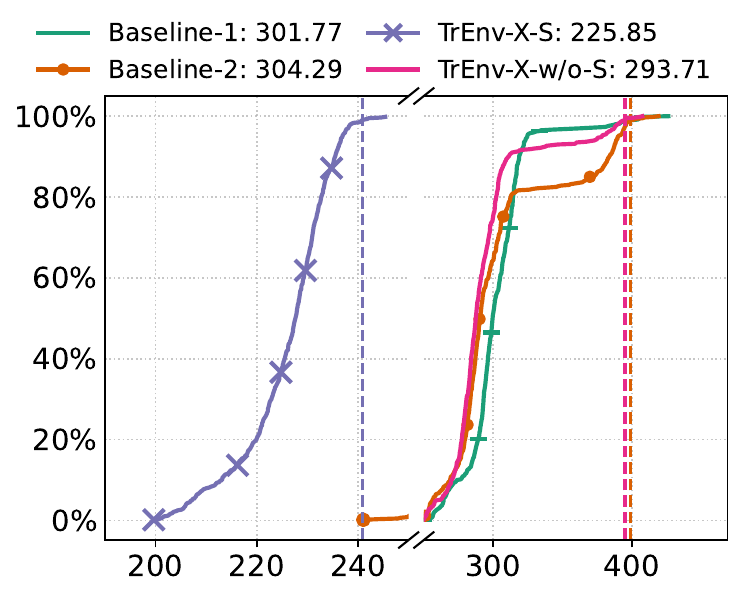}
        \subcaption{Blog summary.}
    \end{minipage}
    \hfill
    \begin{minipage}[t]{0.33\textwidth}
        \centering
        \includegraphics[width=\linewidth]{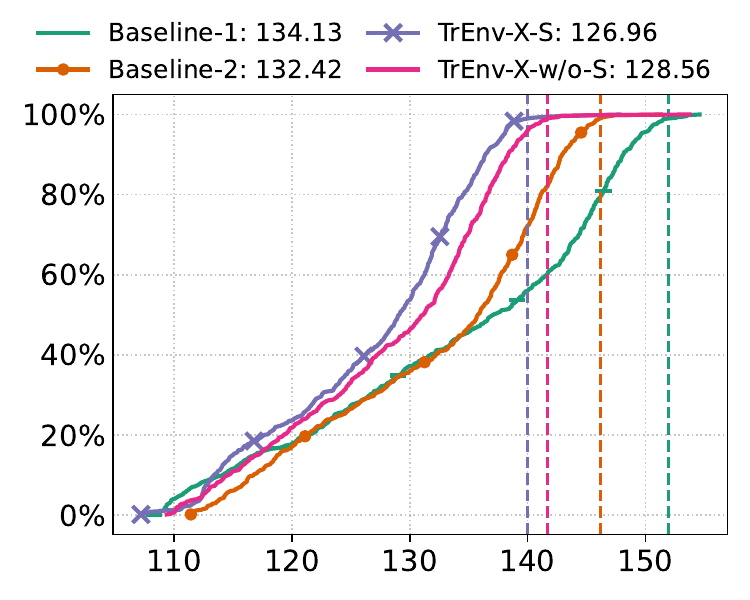}
        \subcaption{Game design.}
    \end{minipage}    
    \caption{The CDF of different agents' E2E latency. The x-axis denotes the E2E latency in seconds. The vertical dotted line indicates the P99 E2E latency, while numbers in the legend denotes average E2E latency.}
    \label{fig:eval-browser-share}
\end{figure}

As described in \S\ref{sec:browser-share}, our first optimization, browser sharing, is designed to mitigate performance degradation under CPU overcommitment. We evaluate this technique using the last three agents listed in Table \ref{tab:agent-chara}, as only these agents utilize a browser during execution.
Figure \ref{fig:eval-browser-share} presents the results, where TrEnv-S denotes the \systemname system augmented with browser sharing.
Browser sharing reduces the P99 and average end-to-end latencies by 2\%–58\% and 1\%–26\%, respectively, depending on the agent. The performance gains vary significantly across agents. For instance, the Game Design agent shows minimal improvement compared to the other two. As shown in Table \ref{tab:agent-chara}, this agent has a low CPU utilization of approximately 6\%, indicating infrequent browser usage. As a result, the benefits of browser sharing have limited impact on its latency.
In contrast, agents with more frequent browser interactions, such as Blog Summary, experience substantial performance improvements. In this case, browser sharing reduces the P99 latency by up to 58\%, demonstrating the effectiveness of this optimization for browser-intensive workloads.

\subsubsection{Mitigation of duplicated page cache.}\label{sec:eval-mitigate-page-cache-agents}
\begin{figure}[t!]
    \centering
    \includegraphics[width=0.65\linewidth]{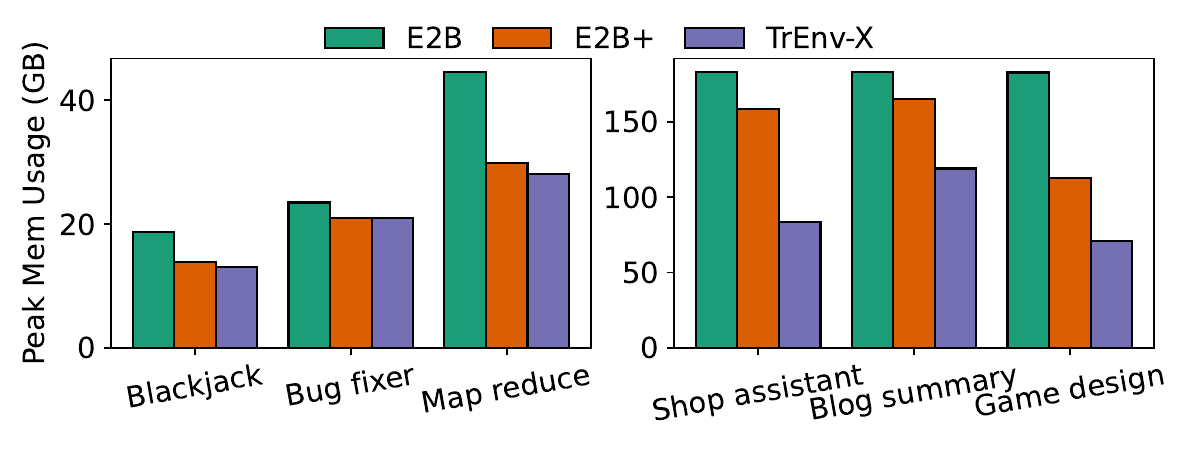}
    \caption{The peak memory usages of different agents.}
    \label{fig:agent-mem-use}
\end{figure}

\begin{figure}[t!]
    \centering
    \includegraphics[width=0.7\linewidth]{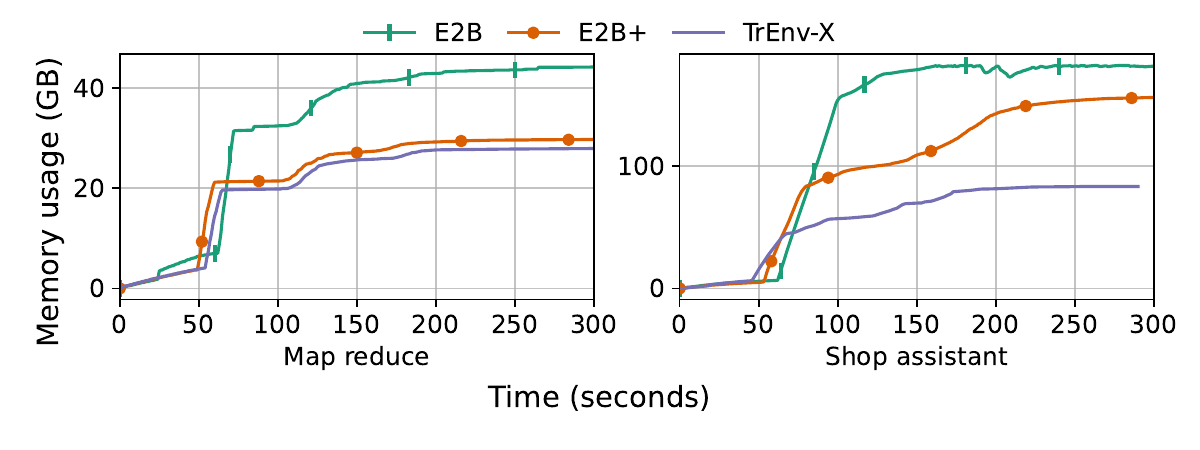}
    \caption{The memory usage during execution of Map reduce and Blog summary agent.}
    \label{fig:agent-mem-time-series}
\end{figure}

To address the issue of duplicated page cache, \systemname leverages the virtio-pmem device in combination with a union filesystem. This design allows multiple VMs to share common file contents in memory, reducing redundant caching across instances.
As shown in Figure \ref{fig:agent-mem-use}, \systemname reduces peak memory usage by approximately 10\%–61\% compared to \agentB, and up to 48\% compared to \agentBp. The extent of memory savings varies across agents, primarily depending on the proportion of file access during execution. For instance, agents such as Blackjack and Bug Fixer perform minimal file I/O, resulting in limited memory optimization benefits.
In addition to peak memory usage, we also track memory utilization throughout the agent’s execution, as illustrated in Figure \ref{fig:agent-mem-time-series}. If memory cost is modeled as the product of usage and duration, \systemname achieves over 50\% savings in overall memory cost, highlighting its effectiveness in reducing resource consumption across time.

}

%% file: content/related-work.tex
\section{Related Work}\label{sec:related-work}


\noindent \textbf{Sandboxing techniques}. Prior research has explored various sandboxing techniques like microVMs~\cite{agache_firecracker_nodate,gvisor2025,zijun_rund_2022}, lightweight containers~\cite{oakes_sock_nodate}, unikernels~\cite{cadden_seuss_2020}, and WebAssembly-based sandboxes~\cite{gadepalli_sledge_2020,zhang_narrowing_2019,shillaker_faasm_nodate}.
The current implementation of \systemname primarily focuses on containers and microVMs.

\noindent \textbf{Caching and pre-warm.}
Prior studies have sought to reduce the overhead of cold starts by employing caching or pre-warming techniques. 
After execution, the platform may keep the container alive for a certain duration, reusing it if the same function is invoked again within that period.
Besides, the platform may pre-start containers for functions that are likely to be executed soon.
Some platforms utilize fixed keep-alive times or static pre-warm strategies~\cite{openwhisk-warmed-container,wang_peeking_nodate}, while recent research has explored heuristic approaches~\cite{shahrad_serverless_nodate,roy_icebreaker_2022,fuerst_faascache_2021}, including the use of machine learning models~\cite{xu_adaptive_2019}.
\systemname takes a different approach by directly reducing cold start overhead, thereby eliminating the need for designing those complex strategies.
Groundhog~\cite{alzayat_groundhog_2023}, a lightweight sequential request isolation system, restores memory to a ``clean'' state before reuse in its caching strategy.
In contrast, \systemname focuses on mitigating cold start overhead by sharing resources across different functions and hosts.
Additionally, \systemname is capable of providing functionality similar to Groundhog by fully restoring the container state after each execution.

\noindent \textbf{Utilization of ``fork''}. 
Several studies have investigated the use of the OS `fork' primitive for efficiently initiating new instances from a cached container or zygote~\cite{du_serverless_2022,du_catalyzer_2020,li_help_nodate,oakes_sock_nodate}. 
Moreover, MITOSIS~\cite{wei_no_2022} enhances the OS by implementing a remote fork primitive that enables the use of states from remote machines.
While \systemname shares some similarities with MITOSIS, there are significant differences.

Firstly, MITOSIS utilizes less-isolated, lightweight containers to reduce isolation costs, which results in compromised isolation levels, as discussed in Section \ref{sec:rootfs-reconfig} and \ref{sec:reused-kernel-objects}. 
Secondly, MITOSIS is currently limited to supporting single-thread functions. Overcoming this limitation typically requires specialized solutions that involve both the language runtime and the OS levels, complicating the development of a universal ``fork'' solution. 
In contrast, \systemname is built upon CRIU, which supports the restoration of multi-threaded and multi-process environments more robustly.
Although MITOSIS achieves startup latencies of less than 4 ms for many single-threaded functions, initializing real-world serverless functions that involve multiple processes and threads is inherently more complex and costly. 
Furthermore, while both MITOSIS and \systemname utilize kernel-space RDMA, MITOSIS depends on a single container (i.e., the parent) to fork from, thus focusing primarily on optimizing RDMA's scalability.
In contrast, \systemname leverages disaggregated memory, focusing on how memory is transparently shared. 
Moreover, \systemname is designed to accommodate various types of disaggregated memory technologies, including RDMA and CXL.

\noindent \textbf{Storage Systems.}
Many researchers have tried to optimize serverless computing by enabling more efficient state transfer~\cite{romero_faat_2021,sreekanti_cloudburst_2020,wang_infinicache_nodate}. They build scalable and distributed storage or cache systems.
These works are orthogonal to \systemname and can be integrated to enhance \systemname's I/O performance.

%% file: content/conclusion.tex
\section{Conclusion}

We presented \systemname, a serverless platform that achieves extreme elasticity by sharing resources. It reduces isolation overhead through repurposable sandboxes and enables fast memory restoration via mm-templates.
\newc{
Evaluating LLM-based agents, a representative serverless workload, observe that their execution cost on existing serverless platforms is comparable to that of LLM inference.
To address this, \systemname is extended to support VM-based agents, and introduce optimizations such as browser sharing and DAX, which enable fast startup and high-density deployment.
}
In our evaluation, \systemname in a container-based setup reduces memory usage by 48\% and improves P99 latency by up to 7$\times$, when compared against prior systems like FaaSnap and REAP.
\newc{
For VM-based scenarios, \systemname reduces agents' memory usage by up to 61\% and P99 latency by 58\% compared to E2B.
}